\documentclass[preprint,nofootinbib,aps,superscriptaddress,eqsecnum]{revtex4-1} 
 \pdfoutput=1
 \usepackage{graphicx}
 \usepackage{amsmath}
\usepackage{amsfonts}
\usepackage{amssymb}
\usepackage{caption}
\usepackage{subcaption}
\def\bea{\begin{eqnarray}}
\def\eea{\end{eqnarray}}
 \def\be{\begin{equation}}
\def\ee{\end{equation}}

\begin{document}
  
\title{ Electroweak vacuum stability in presence of singlet scalar dark matter in TeV scale seesaw models }

 \author{Ila Garg}
\email[Email Address: ]{ila.garg@iitb.ac.in}
\affiliation{Department of Physics, Indian Institute of Technology Bombay, Powai, Mumbai 400 076, India}
\affiliation{Theoretical Physics Division, 
Physical Research Laboratory, Ahmedabad - 380009, India}

\author{Srubabati Goswami}
\email[Email Address: ]{sruba@prl.res.in}
\affiliation{Theoretical Physics Division, 
Physical Research Laboratory, Ahmedabad - 380009, India}

 \author{Vishnudath K. N.}
\email[Email Address: ]{vishnudath@prl.res.in}
\affiliation{Theoretical Physics Division, 
Physical Research Laboratory, Ahmedabad - 380009, India}
\affiliation{Discipline of Physics, Indian Institute of Technology, Gandhinagar - 382355, India}

\author{Najimuddin Khan}
\email[Email Address: ]{najimuddin@prl.res.in}
\affiliation{Theoretical Physics Division, 
Physical Research Laboratory, Ahmedabad - 380009, India}

\begin{abstract}
We consider singlet extensions of the standard model, both in the fermion 
and the scalar sector,  to account for 
the generation of neutrino mass at the TeV scale and the existence of dark matter respectively. 
For the neutrino sector we consider models with extra singlet fermions which 
can generate neutrino mass via the so called inverse or linear seesaw 
mechanism whereas a singlet scalar is introduced as the candidate for
dark matter.
We show that although  these
two sectors are disconnected at low energy, 
the coupling constants of both the sectors get  correlated  
at high energy scale by the constraints coming from the  perturbativity and stability/metastability of the electroweak vacuum. 
The singlet fermions try to destabilize the electroweak vacuum 
while the singlet scalar aids the stability. As an upshot, the electroweak vacuum may attain absolute stability 
even upto the Planck scale for suitable values of the parameters.
We delineate the parameter space for the singlet 
fermion and the scalar couplings for which the electroweak vacuum remains 
stable/metastable and at the same time giving the correct relic density and neutrino 
masses and mixing angles as observed.     

 \end{abstract}
 
\maketitle
 \section{Introduction}

The Large Hadron Collider (LHC) experiment has completed the hunt for the last missing piece of the 
Standard Model (SM)  with the discovery of the Higgs boson
\cite{Chatrchyan:2012xdj,Aad:2012tfa}. 
The Higgs boson holds a special status in the SM as it gives mass to 
all the other particles, with the exception of the neutrino. 
However, observation of neutrino oscillation, from solar, atmospheric, reactor
and accelerator experiments necessitates  the extension of the SM to incorporate 
small neutrino masses.    
The seesaw mechanism is considered to be the most elegant way to 
generate small neutrino masses. 
The origin of seesaw is from the dimension 5 effective  operator
$\kappa LLHH $, proposed by Weinberg in \cite{weinberg}.  
Here, $L$ and $H$ are the SM lepton, and Higgs fields
respectively. $\kappa$ is a coupling constant with inverse mass dimension.  
This term violates lepton number by two units 
and imply that neutrinos are Majorana particles. 
The generation of the effective dimension 5 operator needs 
extension of the SM by new particles.
The most minimal scenario in this respect 
is the canonical type-1 seesaw model, 
in which the SM is extended by heavy right handed Majorana neutrinos 
for ultra-violet completion of the theory \cite{Minkowski:1977sc,seesaw1,seesaw2,Mohapatra:1979ia}.
The essence of seesaw mechanism lies in the fact that the lepton
number is explicitly violated at a high-energy scale which defines the 
scale of the new physics.  
However to give an observed neutrino mass of the order of $m_\nu \sim 0.01 eV$ 
one needs the Majorana neutrinos to be very heavy ($\sim 10^{15}$ GeV), 
close to the scale of Grand Unification. 
However since such high scales are not accessible to colliders, in the 
context of the LHC, there have been a proliferation of 
studies involving TeV scale seesaw models. For recent reviews see for 
instance \cite{Boucenna:2014zba,Deppisch:2015qwa}.  
For ordinary seesaw mechanism lowering the scale of 
new physics to  TeV requires small Yukawa couplings 
${\cal{O}}(10^{-6})$ \footnote{Unless very special textures leading to cancellations are invoked 
\cite{Adhikari:2010yt,Kersten:2007vk,Pilaftsis:1991ug}.} 
and for such values, the light-heavy mixing is small 
and no  interesting collider signals can be studied. 
One of the ways to reduce the scale of new physics to TeV is to 
decouple the new physics scale from the scale of lepton number violation. 
The smallness of the neutrino mass can then be attributed to small lepton 
number violating terms. A tiny value  of the latter is deemed natural,  
since when this parameter is zero, the global U(1) lepton number symmetry
is reinstated  and neutrinos are massless. 
One of the most popular TeV scale seesaw models based on the above 
idea  is the 
inverse seesaw model
\cite{Mohapatra:1986bd}.  This contains additional singlet states ($\nu_s$), 
along with the right handed neutrinos ($N_R$), having opposite lepton numbers. 
The lepton number is broken softly by introducing a small 
Majorana mass term for  the singlets. This parameter is responsible for 
the smallness of the neutrino mass and one does not 
require small Yukawa couplings to get observed neutrino masses 
and at the same time the scale of new physics can be at TeV.
Another possibility of a TeV scale singlet seesaw model is the linear seesaw model 
\cite{Gu:2010xc,Zhang:2009ac,Hirsch:2009mx}. 
The difference is, in this case, a small lepton number violating term 
is generated by the
coupling between the the left-handed neutrinos and the singlets states.
The inverse seesaw and linear seesaw differ from each other in the
way lepton number violation is introduced in the model, as we will see in 
the next section. Also, the particle content of the minimal models that 
agree with the oscillation data for these two are different. 
For linear seesaw, we need only one $N_R$ and one $\nu_s$
\cite{Gavela:2009cd,Khan:2012zw,Bambhaniya:2014kga}
 whereas in the inverse seesaw case, we need two $N_R$ and two $\nu_s$ \cite{Malinsky:2009df}.
Note that the minimal linear seesaw model is the simplest re-constructable Tev scale seesaw model having a
minimum number of independent parameters.

Apart from neutrino mass, another issue that requires extension of 
the SM is the existence of dark matter. 
Measurements  by  Planck  and  WMAP  demonstrate  that  nearly  
85 percent of  the  Universe's 
matter density is dark \cite{Ade:2015xua}. 
Among the various models of dark matter that are proposed in the literature, 
the most minimal renormalizable extension of the SM 
are the so called Higgs portal models
\cite{ Silveira:1985rk,McDonald:1993ex,Burgess:2000yq}. 
These models include a scalar singlet  
that couples only to the Higgs. 
An additional $Z_2$ symmetry  
is imposed to prevent the decay of the DM and safe-guard its
stability. The coupling of the singlet with 
the Higgs provides the only  portal 
for its interaction with the SM. 
Nevertheless  
there can be testable consequences of this scenario which can 
put constraints on its coupling and  mass. 
These include constraints from searches of 
invisible decay of Higgs at the Large Hadron Collider (LHC)
\cite{Belanger:2013xza,Aad:2015pla,Khachatryan:2016whc}, 
direct and indirect detections of DM as well as compliance with 
the observed relic density \cite{ Dick:2008ah,Yaguna:2008hd,Cai:2011kb,Urbano:2014hda,Cuoco:2016jqt}. 
Implications for 
the LHC \cite{Barger:2007im,Cheung:2015dta,Djouadi:2011aa,Endo:2014cca, Han:2016gyy} and ILC \cite{Ko:2016xwd} have also been studied.
Combined constraints from all these have been discussed in
\cite{Mambrini:2011ik,Cheung:2012xb,Cline:2013gha} and most recently in 
\cite{Athron:2017kgt}. 

The singlet Higgs can also affect the stability of the 
electroweak vacuum
\cite{Gonderinger:2009jp, EliasMiro:2012ay, Chen:2012faa, Khan:2014kba, Haba:2013lga, Ghosh:2015apa}. 
It is well known that the electroweak vacuum in the standard model is 
metastable and the Higgs quartic coupling $\lambda$ is 
pulled down to negative value by renormalization group running, 
at an energy of about $10^{9}-10^{10}$ GeV, 
depending on the value of $\alpha_s$ and the top quark mass $m_t$, 
as the dominant contribution comes from the top-Yukawa coupling, 
$y_t$ \cite{Alekhin:2012py,Buttazzo:2013uya}. 
This indicates the existence of another low lying vacuum.  
If the quartic coupling $\lambda(\mu)$ becomes negative at large 
renormalization scale $\mu$, it
implies that in the early universe the Higgs potential would be unbounded from 
below and the vacuum would be unstable in that era. 
But it does not pose any threat to the standard model as it has been shown that the decay time is greater than the age of the universe \cite{Isidori:2001bm}.
In the context of standard model extended with neutrino masses via canonical type-1 seesaw mechanism,
the Yukawa coupling of the RH neutrinos also contribute to the RG running, 
just like $y_t$ and thereby we expect it to affect the electroweak vacuum stability negatively. But this effect is not so much because, as discussed before,
in order to get the
light neutrino masses, either one has to resort to extremely small
Yukawa couplings or one needs a very large Majorana mass scale $(\approx 10^{15}\,\textrm{GeV})$ and the contribution to the running of $\lambda$ is
much smaller in both the cases compared to that from $y_t$.
However, for the 
TeV scale seesaw models, with sizable Yukawa 
couplings the stability of the vacuum  can be altered considerably  
by the contribution from the neutrinos
\cite{Rodejohann2012,Chakrabortty:2012np,Khan:2012zw,Datta:2013mta,Kobakhidze:2013pya,
Rose:2015fua,Bambhaniya:2016rbb,Bambhaniya:2014hla,Lindner:2015qva}. 
On the other hand, the  
singlet scalar can help in stabilizing the 
electroweak vacuum  by adding a positive contribution which 
prevents the Higgs quartic coupling from becoming negative.  The stability of the electroweak vacuum in the context of singlet scalar extended SM with an unbroken $Z_2$ symmetry has been explored 
 in \cite{Gonderinger:2009jp,Chen:2012faa,Haba:2013lga,Khan:2014kba}.

In this paper, we extend the SM by adding extra fermion as well as 
scalar singlets to explain the origin of neutrino mass as well 
as existence of  dark matter \footnote{ For other studies to explain 
neutrino mass and dark matter using scalar singlets 
see for instance \cite{Davoudiasl:2004be,Bhattacharya:2016qsg}.}. 
The candidate for dark matter is a real singlet scalar added to 
SM with an additional $Z_2$ symmetry which ensures its stability. 
For generation of neutrino mass at TeV scale we consider two 
models. The first one is the general inverse seesaw model 
with three right handed neutrinos and three additional singlets. 
The second one is the minimal linear seesaw model.
These two sectors are disconnected at the low energy. 
However, the consideration of the stability of the electroweak vacuum and perturbativity
induces a correlation between the two sectors. 
We study the stability of the electroweak vacuum in this model 
and explore the effect of the two opposing trends -- singlet fermions 
trying to destabilize the vacuum further 
and singlet Higgs trying to oppose this. 
We find the parameter space, which is consistent 
with the constraints of relic density and neutrino oscillation 
data and at the same time can cure the instability of the  electroweak 
vacuum. We present some benchmark points 
for which the electroweak vacuum is stable up to the Planck's scale. 
In addition to absolute stability we also explore the parameter region 
which gives metastability in the context of this model. 
We investigate the combined effect of these two sectors and  obtain the 
allowed parameter space consistent with observations and vacuum stability/metastability and perturbativity.  

The plan of the paper is as follows. In the next section we discuss the 
TeV scale singlet seesaw models, in particular the inverse seesaw 
and linear seesaw mechanism. We also outline the diagonalization procedure
to give the low energy neutrino mass matrix. In section III we discuss   
the potential in presence of a singlet scalar. 
Section IV presents the effective Higgs potential and the renormalization group 
(RG) evolution of the different couplings. In particular we include the contribution from both fermion and scalar singlets in the effective potential. 
In section V we discuss the existing constraints on the fermion and 
the scalar sector couplings from experimental observations and also from perturbativity. 
We present the results in section VI and conclusions in section VII.  .

\section{TeV scale Singlet Seesaw Models}
 
The most general low scale singlet seesaw scenario consists of adding $m$ right handed neutrinos $N_R$ and $n$ gauge-singlet sterile neutrinos  $\nu_s $ to the standard model. 
   The lepton number for $\nu_s$ is chosen to be $-1$ and that for $N_R$ is $+1$. For simplicity, we will  work in a basis where the charged 
   leptons are identified with their mass eigenstates. We can write the most general Yukawa part of the Lagrangian 
   responsible for neutrino masses, before spontaneous symmetry breaking(SSB) as,\be
 -L_{\nu} = \overline{l}_LY_\nu\,{H}^cN_R \,\,+ \,\,\overline{l}_LY_s\,{H}^c\nu_s \,\,+\overline{N_R^c} \, M_R\,\nu_s  \,\,+\,\, \frac{1}{2}\,\overline{\nu_s^c}M_\mu \nu_s \,\,\, +\,\, \frac{1}{2}\,\overline{N_R^c}M_N N_R \,\,\, +\, \textrm{h.c.} \ee
where $l_L$ and $H$ are the lepton and the Higgs doublets respectively,
 $Y_\nu$ and $Y_s$ are the Yukawa coupling matrices, $M_N$ and $M_\mu$ are the symmetric Majorana mass matrices for $N_R$ and $\nu_s$ respectively. $Y_\nu$, $Y_s$, $M_N$ and $M_\mu$ are of dimensions $3 \times m, \,3 \times n,\, m \times m \,$ and $n \times n$ respectively.  
 
 Now, after symmetry breaking, the above equation gives,
  \be
 -L_{mass} = \overline{\nu}_L M_DN_R \,\,+ \,\,\overline{\nu}_L M_s\nu_s \,\,+\overline{N_R^c} \,M_R \nu_s \,\,+\,\, \frac{1}{2}\,\overline{\nu^c_s}M_\mu \nu_s \,\,+\,\, \frac{1}{2}\,\overline{N_R^c} M_N N_R\,\,\, +\,\,\, \textrm{h.c.} \ee
 
 where, $M_D = Y_\nu \langle H \rangle \,  $ and $M_s = Y_s \langle H \rangle \,$. 
 The neutral fermion mass matrix $M$ can be defined as,
 \be -L_{mass} = \,\,\ \frac{1}{2}( \, \overline{\nu}_L \,\, \, \overline{N_R^c}\,\,\, \overline{\nu^c_s}\,) \begin{pmatrix}
  0 & M_D & M_s \\
 M_D^T & M_N & M_R \\
 M^T_s & M_R^T & M_\mu
\end{pmatrix}  \begin{pmatrix}
 \nu_L^c \\
 N_R \\
 \nu_s
\end{pmatrix}  \,\,\, + \,\,\,\textrm{h.c.} \label{gen}\ee
 
 From this equation, we can get the variants of the singlet  seesaw scenarios by setting certain terms to be zero.

 \subsection{Inverse Seesaw Model (ISM)}
 
In the inverse seesaw model, $M_s$ and $M_N$ are taken to be zero \cite{Mohapatra:1986bd}. The mass scales of the three 
sub-matrices of $M$ may naturally have a hierarchy  $ \, M_R >>  M_D >> M_\mu \,$, 
 because the mass term $M_R$ is not subject to the $SU(2)_L$ symmetry breaking and the mass term $M_\mu$ violates the lepton number. 
 Thus we can take $M_\mu$ to be naturally small by t' Hooft's naturalness criteria since the expected degree of lepton number violation in nature is very small. 
 In this paper, we consider a (3+3+3) scenario for the inverse seesaw model for generality and hence all the three sub-matrices $M_R$, $M_D$ and $M_\mu$ are $3 \times 3$ matrices.
  The effective light neutrino mass matrix in the seesaw approximation is  given by, 
\be M_{light} \,\, = \,\, M_D (M_R^T)^{-1}M_\mu M_R^{-1} M_D^T \label{eqISMmlight}  \ee
 and in the heavy sector, we will have three pairs of degenerate pseudo-Dirac neutrinos of masses of the order  $\sim$ $M_R \, \pm \, M_\mu$. 
Note that the smallness of $M_{light}$ is naturally attributed to the smallness of both $M_\mu$ and $ \,\, \frac{M_D}{M_R}$.  
For instance, $M_{light} \sim \mathcal{O}\,(0.1)$ eV can easily be achieved for  $ \,\, \frac{M_D}{M_R} \sim 10^{-2}\,\,$ 
and $\,M_\mu \sim \mathcal{O}\,$(1) keV. Thus, the seesaw scale can be lowered down considerably assuming $Y_\nu \,\sim \, \mathcal{O}(0.1)$, such that $M_D \,\sim \, 10 $ GeV and $M_R \, \sim \, 1$ TeV.
 
  \subsection{Minimal Linear Seesaw Model (MLSM)}\label{minimal}
  In eqn. (\ref{gen}), if we put $M_N$ and $M_\mu$ to be zero and choose the 
  hierarchy $M_R$  $>>$ $M_D >> M_\mu$, we will get the linear seesaw model \cite{Gu:2010xc,Zhang:2009ac,Hirsch:2009mx}.
  In this paper, we consider the minimal linear seesaw model in which we add only one
  right handed neutrino $N_R$ and one gauge-singlet sterile neutrino  $\nu_s $ \cite{Gavela:2009cd,Khan:2012zw,Bambhaniya:2014kga}.
  In such a case, the lightest neutrino mass is zero.  
  The source of lepton number 
  violation is through the coupling $Y_s$  which is assumed to be very small. 
  Here, $Y_\nu$ and $Y_s$ are the $(3\times1)$ Yukawa coupling matrices and the overall neutrino mass matrix is a symmetric matrix of dimensions $5 \times 5$.
  The light neutrino mass matrix to the leading order is given by,
\be  
 M_{light}\,=\,M_D (M_R^T)^{-1} M_S^T\,+\,M_S (M_R^T)^{-1}{M_D^T}.
\ee 
  Assuming $M_D\,\sim\, 100$ GeV and $M_R\,\sim \,1$ TeV, 
 one needs $Y_s\,\sim \, 10^{-11}$ to get light neutrino mass $m_\nu\, \sim \,0.1$ eV.
 The heavy neutrino sector will consist of a pair of degenerate neutrinos.

  \subsection{Diagonalization of the Seesaw Matrix and Non-unitary PMNS Matrix}

The diagonalization procedure is same for both the cases. Here we illustrate it for the inverse seesaw case.
The $9 \, \times 9$ inverse seesaw  mass matrix can be rewritten as,
  \be M_\nu \, = \, \begin{pmatrix}
 0 & \hat{M}_D \\
 \hat{M}_D^T & \hat{M}_R
\end{pmatrix} \label{matmnu} \ee
  where, $\, \hat{M}_D = \, (M_D \,\,\,\, 0) \,\,$ and $\, \, \hat{M}_R = \, \begin{pmatrix}
 0 & M_R \\
 M_R^T & M_\mu
\end{pmatrix} $. 
We can diagonalize the neutrino mass matrix
using a $9 \times 9$ unitary matrix \cite{Xing:2005kh,Grimus:2000vj},
\be U_0^T \,M_\nu \, U \,\, = \,\, M_\nu^{diag}  \ee
 where, $M_\nu^{diag} \, = \,\textrm{ diag}\,(m_1 ,\, m_2 ,\, m_3 ,\, M_1 ,\,..., \,M_6 )\,$ with mass eigenvalues $\,m_i \, (i =
1 ,\, 2 ,\, 3)\,$ and $\,M_j\, (j = 1 , \,... , 6)\,$ for three light neutrinos and 6 heavy neutrinos respectively. 
  Following the two-step diagonalization procedure, $U_0$ could be expressed as, (by keeping terms up to order $\mathcal{O}(\hat{M}_D^2/\hat{M}_R^2)$) \cite{Grimus:2000vj}
\be U_0 =  W\,T =\, \begin{pmatrix}
 U_L & V \\
S & U_H
\end{pmatrix} \,=\, 
\begin{pmatrix}
 (1-\frac{1}{2}\epsilon)U_\nu & \hat{M}_D^* (\hat{M}_R^{-1})^*U_R \\
-\hat{M}_R^{-1}\hat{M}_D^T \, U_\nu & (1-\frac{1}{2}\epsilon ')U_R
\end{pmatrix}. \label{UL2}
\ee
Here, $\, U_L,\, V,\,S\,\, \textrm{and} \,\,U_H\,$ are $3 \times 3 \,,\, 3 \times 6 \,,\, 6\times 3 \,\, \, \textrm{and}\,\, 6\times 6\, $ 
matrices respectively, which are not unitary.
W is the matrix which brings the full $9 \times 9$ neutrino matrix, in the
block diagonal form,
\be W^T \begin{pmatrix}
 0 & \hat{M}_D \\
 \hat{M}_D^T & \hat{M}_R
\end{pmatrix} W \,=\,   \begin{pmatrix}
 M_{light} & 0 \\
 0 & M_{heavy}
\end{pmatrix}  ,   \ee
$T \, = \, \textrm{diag}\,(U_\nu , U_R ) \,  $ diagonalizes the mass matrices in the light and heavy sectors
appearing in the upper and lower block of the block diagonal matrix respectively. In the seesaw limit,
$M_{light}$ is given by eqn. (\ref{eqISMmlight})  and $ M_{heavy}$ = $\hat{M}_{R}$. In eqn. (\ref{UL2}), $U_L$ corresponds to the PMNS matrix which acquires a non-unitary 
correction $(1-\frac{\epsilon}{2})$. The parameters $\epsilon$ and $\epsilon '$ characterize the non-unitarity and are given by,
\be \epsilon \, = \, \hat{M}_D^* \hat{M}_R^{-1*}\hat{M}_R^{-1}\hat{M}_D^T \, , \ee
\be \epsilon ' \, = \, \hat{M}_R^{-1}\hat{M}_D^T \hat{M}_D^* \hat{M}_R^{-1*}. \ee

\section{Scalar Potential of the Model}

As mentioned earlier, in addition to the extra fermions, we also add an extra real scalar singlet $S $ to the standard model.
The potential for the scalar sector with an extra $Z_2$ symmetry under $S\, \rightarrow \, -S\,$ is given by,
\be  V(S , H)\,\, = \,\, {m^2}H^\dagger H \,+\, \
{\lambda}(H^\dagger H)^2 \,+\, \frac{\kappa}{2} H^\dagger H \,S^2 \,+ \, \frac{m_S^2}{2}S^2 \,
+\,\frac{\lambda_{S}}{24}S^4 \, \label{eqpot1}. \ee

In this model, we take the vacuum expectation value ($vev$) of $ S$ as $0$, so that $Z_2$ symmetry is not broken. The standard model scalar doublet $H$  could be written as,
\be   H = \frac{1}{\sqrt{2}}\begin{pmatrix}
  G^{+}\\
 v+h+iG^0
\end{pmatrix}   \ee
 where the $vev$ $v = 246 ~\textrm{GeV}$.

Thus, the scalar sector consists of two  particles $h$ and $S$, where $h$ is the standard model Higgs boson with a mass of $\sim$ $126 \, GeV$, and  the mass of  the extra scalar is  given by,
 \be M_{DM}^2 \,=\, {m_S}^2\,+\, \frac{\kappa}{2}v^2.   \ee

As the $Z_2$ symmetry is unbroken upto the Planck scale $M_{pl} = 1.22 \times 10^{19}$ GeV, the potential can have minima only along the Higgs field direction and also this symmetry prevents the extra scalar from acquiring a vacuum expectation value. This extra scalar field does not mix with the SM Higgs. Also an $odd$ number of this extra scalar does not couple to the standard model particles and the new fermions. As a result, this scalar is stable and serve as a viable weakly interacting massive dark matter particle. 
The scalar field $S$ can annihilate to the SM particles as well as to the new fermions only via the Higgs exchange. So it is called a Higgs portal dark matter. 

\section{Effective Higgs Potential and RG evolution of the Couplings}

 The effective Higgs potential and the renormalization group equations  are the same for both  the linear and the inverse seesaw models. The two models differ only by the 
 way in which a small lepton number violation is introduced in them, whose effect could be neglected in the RG evolution. So, effectively, the RGEs are the same in both the models, the only difference
 being the dimensions of the Yukawa coupling matrices and the number of heavy neutrinos present in the model.

\subsection{Effective Higgs Potential}

The tree level Higgs potential in the standard model is given by, 
\be V(H) \,= \, -m^2 H^\dag H \, +\, \lambda (H^\dag H)^2 .  \ee
This will get corrections from higher order loop diagrams of SM particles. In the presence of the extra singlets, 
the effective potential will get additional contributions from the extra scalar and the fermions. Thus, we have the one-loop effective Higgs potential $(V_1(h))$ in our model as,
\be V_1^{SM+S +\nu}(h)  \, = \, V_1^{SM}(h) \, + \, V_1^{S}(h) \, + \, V_1^{\nu}(h)  \ee
where the one loop contribution to the effective potential due to the standard model particles is given by \cite{Casas:1994qy,Casas:1996aq},

\be  V_1^{SM}(h) \, = \, \sum_{i}\frac{n_i}{64 \pi^2} M_i^4(h) \Bigg[ \textrm{ln} \,\frac{M_i^2(h)}{\mu^2(t) }\, - \, c_i  \Bigg]  \label{4.4}.\ee

Here, the index $i$ is summed over all SM particles and $c_{H,G,f} \, = \, 3/2$ and $c_{W,Z}\,=\, 5/6$, where $H,\, G,\, f,\, W$ and $Z$ stand for the Higgs boson, the Goldstone boson, fermions and $W$ and $Z$ bosons respectively ; $M_i(h)$ can be expressed as,
$$M_i^2(h) \,=\,\kappa_i(t)\,h^2(t) \,-\, \kappa'_i(t).$$ 

 The values of  $n_i$, $\kappa_i$ and $\kappa_i'$ are given in the eqn. (4) in \cite{Casas:1994qy}. Here $ h\,= \,h(t)$ denotes the classical value of the Higgs field, $t$ being the dimensionless parameter related to the running energy scale $\mu$ as $t~ =~ log(\mu/M_{Z})$.
 
The one loop contribution due to the extra scalar is given by \cite{Lerner:2009xg,Gonderinger:2012rd}
\be  V_1^{S}(h) \, = \, \frac{1}{64 \pi^2} M_S^4(h) \Bigg[ \textrm{ln} \,\frac{M_S^2(h)}{\mu^2(t)}  \, - \, \frac{3}{2} \Bigg]  .\ee
where
$$M_S^2(h) \,=\,m_S^2(t) \,+\, \kappa(t) h^2(t)/2$$

The contribution of the extra neutrino Yukawa coupling to the one loop effective potential can be written as \cite{Casas:1999cd,Khan:2012zw},
\be V_1^{\nu} (h) = -\frac{((M'^\dag M')_{ii})^2}{32 \pi^2} \Bigg[ \textrm{ln}\,\frac{(M'^\dag M')_{ii}}{\mu^2(t)}  -\frac{3}{2}\Bigg]-\frac{((M' M'^{\dag})_{jj})^2}{32 \pi^2} \Bigg[ \textrm{ln}\,\frac{(M' M'^\dag)_{jj}}{\mu^2(t)}  -\frac{3}{2}\Bigg].
 \ee
 
Here $M' = M_{D}$ for inverse seesaw and $M' = ( M_{D} \,\,\, M_{s} )$ for linear seesaw. Also, $i$ and $j$ run over three light neutrinos and $m$  heavy neutrinos to which the light neutrinos are coupled via Yukawa coupling respectively. In our analysis, we have taken two-loop (one-loop) contributions to the effective potential from the standard model particles (extra singlet scalar and fermions).
For $h(t) \, >> \, v$, the effective potential could be approximated as,
\be V_{eff}^{SM+S+ \nu}\, = \, \lambda_{eff}(h) \frac{h^4}{4}  \ee

 with 
  \be  \lambda_{eff}(h) \, = \, \lambda_{eff}^{SM}(h) \, + \, \lambda^{S}_{eff}(h) \, + \, \lambda^{\nu}_{eff}(h)   \ee
where the standard model contribution is,
\be \lambda_{eff}^{SM}(h) \,=\, e^{4 \Gamma (h)}\,[\lambda(\mu = 	h) \,+\, \lambda_{eff}^{(1)}(\mu=h)\,+\,\lambda_{eff}^{(2)}(\mu=h)\, ].  \ee
$\lambda_{eff}^{(1)}$ and $\lambda_{eff}^{(2)}$ are the one- and two- loop contributions respectively and their expressions can be found in \cite{Buttazzo:2013uya}.
The contributions due to the extra scalar and the neutrinos are given by
\be   \lambda_{eff}^{S}(h) \,=\, e^{4 \Gamma (h)}\, \Bigg[  \frac{\kappa^2}{64 \pi^2} \Bigg(   \textrm{ln}\frac{\kappa}{2} - \frac{3}{2}  \Bigg)   \Bigg]  \ee and
\be \lambda^{\nu}_{eff}(h) \, = \, -\frac{e^{4 \Gamma (h)}}{32 \pi^2} \Bigg[  (({Y'}_\nu^\dag {Y'}_\nu)_{ii})^2  \Bigg(   \textrm{ln}\,\frac{({Y'}_\nu^\dag {Y'}_\nu)_{ii}}{2} \,-\frac{3}{2}\,  \Bigg) \, + \,  (({Y'}_\nu {Y'}_\nu^\dag)_{jj})^2  \Bigg(   \textrm{ln}\,\frac{({Y'}_\nu {Y'}_\nu^\dag)_{jj}}{2} \,-\frac{3}{2}\,  \Bigg) \,\Bigg] \label{labmbdaeffnu}\ee
where,
\be \Gamma(h) \, = \, \int_{M_t}^{h} \gamma(\mu)\, d\,\textrm{ln}\,\mu . \ee 

Here $\gamma(\mu)$ is the anomalous dimension of the Higgs field and in eqn. (\ref{labmbdaeffnu}), ${Y'}_\nu = Y_\nu$ for inverse seesaw and ${Y'}_\nu = ( Y_\nu \,\,\, Y_{s} )$ for linear seesaw. The contribution of the singlet scalar to the anomalous dimension is zero \cite{Gonderinger:2009jp} and the contribution from the right handed neutrinos at one loop is given in eqn. (\ref{anom}).

\subsection{Renormalization Group evolution of the couplings from $M_t$ to $M_{planck}$}

We know that the couplings in a quantum field theory get corrections from higher-order loop diagrams and as a result, the couplings run with the renormalization scale. For a coupling $C$, we have the renormalization group equation (RGE), 
\be \mu\frac{dC}{d\mu} \, = \, \sum_{i} \frac{\beta_{C}^{(i)}}{(16\pi^2)^i}  \ee
 where i stands for the $i^{th}$ loop.

We have evaluated the SM coupling constants at the  the top quark mass scale and then run them using the RGEs from $m_t$ to $ M_{planck}$. For this, we have taken into account the various threshold corrections at $M_t$ \cite{Sirlin:1985ux,Melnikov:2000qh,Holthausen:2011aa}. All couplings are expressed in terms of the pole masses \cite{Bezrukov:2012sa}.
We have used one-loop RGEs to calculate $g_1(M_t)$ and $g_2(M_t)$ \footnote{Our results are not changed even if we use the two-loop RGEs for $g_1$ and $g_2$.}. For $g_3(M_t)$, we use three-loop RGE  running of $\alpha_s$ where we have neglected the sixth quark contribution and the effect of the top quark has been included using an effective field theory approach. We have also taken the leading term in the four-loop RGE for $\alpha_s$. The mismatch between the top pole mass and the $\overline{MS}$ renormalized coupling has been included. This is given by,
\be  y_t(M_t) \, = \, \frac{\sqrt{2}M_t}{v}\,(1\,+\,\delta_t(M_t))  \ee
where $\delta_t(M_t)$ is the matching correction for $y_t$ at the top pole mass, and similarly for  $\lambda(M_t)$ we have,
\be \lambda(M_t) \, = \, \frac{M_H^2}{2 v^2} \, (1\,+\,\delta_H(M_t)) . \ee
We have included the QCD corrections upto three loops \cite{Melnikov:2000qh}, electroweak corrections upto one-loop \cite{Hempfling:1994ar,Schrempp:1996fb} and the $O(\alpha \alpha_s)$ corrections to the matching of top Yukawa and top pole mass \cite{Bezrukov:2012sa,Jegerlehner:2003py}. Using these corrections, we have reproduced the couplings at $M_t$ as in references \cite{Buttazzo:2013uya,Khan:2014kba}.

Now to evaluate the couplings from $M_t$ to $M_{planck}$, we have used three-loop RGEs for standard model couplings \cite{Buttazzo:2013uya,Mihaila:2012fm,Chetyrkin:2012rz,Zoller:2013mra,Zoller:2012cv}, two-loop RGEs for the extra scalar couplings \cite{Haba:2013lga,Clark:2009dc,Davoudiasl:2004be} and one-loop RGEs for the extra neutrino Yukawa couplings \cite{Antusch:2002rr} \footnote{Our results do not change with the inclusion of two loop RGEs of Neutrino Yukawa couplings which has been checked using SARAH \cite{Staub:2013tta}.}. The one loop RGEs for the scalar quartic couplings and the neutrino Yukawa coupling in our model are given as,
\bea \beta_{\lambda} &=&  \frac{27}{100}g_1^4 +  \frac{9}{10}g_1^2g_2^2  +  \frac{9}{4}g_2^4 -\frac{9}{5}g_1^2\lambda -9g_2^2\lambda +12\lambda^2 +\kappa^2 + 4T\lambda - 4Y  \\&&
\beta_{\kappa} =  -\frac{9}{10} g_1^2\kappa -\frac{9}{2} g_2^2\kappa +6\lambda\kappa +4\kappa^2+ 2T\kappa   \\&&
 \beta_{\lambda_{S}}  = 3\lambda_{S} + 12\kappa^2 \\&&
  \beta_{Y_\nu}  = Y_\nu  \biggr(\frac{3}{2} Y_\nu^\dag Y_\nu - \frac{3}{2}Y_l^\dag Y_l +  T  - \frac{9}{20} g_1^2  -\frac{9}{4}g_2^2  \biggr)\label{RGE}\eea
where, \bea T&=&\textrm{Tr}( 3Y_u^\dag Y_u + 3Y_d^\dag Y_d +  Y_l^\dag Y_l  +  Y_\nu^\dag Y_\nu  ) \nonumber\\&&
Y= \textrm{Tr} (3(Y_u^\dag Y_u)^2 + 3(Y_d^\dag Y_d)^2 + (Y_l^\dag Y_l)^2  +  (Y_\nu^\dag Y_\nu)^2 )  \label{anom}.\eea
 
 The  effect of $\beta$- functions of new particles enters into the SM RGEs at their effective masses.

\section{Exsiting bounds on the fermionic and the scalar sectors}
\label{bounds}
For the vacuum stability analysis, we need to find the Yukawa and scalar couplings that satisfy the existing experimental and theoretical constraints. These bounds are discussed below.
\subsection{Bounds on the fermionic Sector}

\begin{description}

  \item[$\bullet$ Cosmological constraint on the sum of light neutrino masses]  The Planck 2015 results put an upper limit on the sum 
  of active light neutrino masses to be \cite{Ade:2015xua}
   \be\Sigma \, = \, m_1 + m_2 + m_3 < 0.23 \, \textrm{eV} .\ee

  \item[$\bullet$ Constraints from Oscillation data]  
We use the standard parametrization of the PMNS matrix in which,
\be U_\nu = \begin{pmatrix}
  c_{12}c_{13} & s_{12}c_{13} & s13 e^{-i\delta} \\
 -c_{23}s_{12}-s_{23}s_{13}c_{12}e^{i\delta} & c_{23}c_{12}-s_{23}s_{13}s_{12}e^{i\delta} & s_{23}c_{13} \\
 s_{23}s_{12}-c_{23}s_{13}c_{12}e^{i\delta} & -s_{23}c_{12}-c_{23}s_{13}s_{12}e^{i\delta} & c_{23}c_{13}
\end{pmatrix} P \ee
where $ c_{ij} = cos\theta_{ij} \,\, , \,\, s_{ij} = sin\theta_{ij}$ and the phase matrix 
$P = \textrm{diag}\, (1,\,e^{i\alpha_2}, \, e^{i(\alpha_3 + \delta)})$  contains the Majorana phases.
The global analysis \cite{Capozzi:2016rtj,Esteban:2016qun} 
of neutrino oscillation measurements with three 
light active neutrinos give the oscillation parameters in their $3\sigma$ range,  for both normal hierarchy (NH)  for which $m_3 > m_2 >m_1$ and
inverted hierarchy (IH) for which $m_3 > m_2 >m_1$,  as below :

\begin{description}

\item[$\star$ Mass squared differences] 
\begin{equation} \label{osc1} \Delta m_{21}^2/10^{-5} \textrm{eV}^2 \, = \, (7.03 \rightarrow 8.09) \,\,\, ; \,\,\,\,\,
  \begin{cases}
    \Delta m_{31}^2/10^{-3} \textrm{eV}^2 \, = \, (2.407 \rightarrow 2.643) & \text{NH}\\
    \Delta m_{31}^2/10^{-3} \textrm{eV}^2 \, = \, (-2.635 \rightarrow -2.399) & \text{IH}\\
  \end{cases}
\end{equation}

\item[$\star$ Mixing angles] 
   \begin{equation} \label{osc2} \textrm{sin}^2\theta_{12} \, = \, (0.271 \rightarrow 0.345) \,\,\, ;\ee 
   
   \be \label{osc3}  \textrm{sin}^2\theta_{23} =
  \begin{cases}
    (0.385 \rightarrow 0.635) \\
    (0.393 \rightarrow 0.640) \\
  \end{cases}\,\,\,\,\, ; \,\,\,\,\, 
    \textrm{sin}^2\theta_{13} =
  \begin{cases}
    (0.01934 \rightarrow 0.02392) & \text{NH}\\
    (0.01953 \rightarrow 0.02408) & \text{IH}\\
  \end{cases}
\end{equation}
 
 \end{description}

  \item[$\bullet$ Constraints on the non-unitarity of $U_{PMNS}$ = $U_L$]  The analysis of electroweak precision observables 
  along with various other low energy precision observables put bound 
  on the non-unitarity of light neutrino mixing matrix $U_L$ \cite{Antusch:2016brq}. At 90$\%$ confidence level,  
  
  \be |U_L U_L^\dagger|\, =\,  \begin{pmatrix}
  0.9979-0.9998 & <10^{-5} & <0.0021 \\
 <10^{-5} & 0.9996-1.0 & <0.0008 \\
 <0.0021 & <0.0008 & 0.9947-1.0
\end{pmatrix}. \ee
   This also takes care of the constraints coming from various charged lepton flavor violoating decays like $ \, l_i \rightarrow l_j \, \gamma $, among which $\, \, \textrm{Br}(\mu \, \rightarrow e \, \gamma) $ being the one that gives the
most severe bound \cite{TheMEG:2016wtm}, \be \textrm{Br}(\mu \, \rightarrow e \, \gamma)  \,\, < \, 4.2 \times 10^{-13} \ee

\item[$\bullet$ Bounds on the heavy neutrino masses ]  The search for heavy singlet neutrinos at LEP by the
L3 collaboration in the decay channel $N \, \rightarrow \,\, e\,W$ showed no evidence of a singlet neutrino in the mass range between $80 \, \textrm{GeV}  \, (|V_{\alpha i} |^2 \, 
\leq \, 2 \times 10^{-5} )$ and
$205 \, \textrm{GeV} \, (|V_{\alpha i} |^2 \, \leq \,1) $ \cite{Achard:2001qv}, $V_{\alpha i}$ being the mixing matrix elements between the heavy
and light neutrinos. Heavy singlet neutrinos in the mass range from 3 GeV up
 to the Z-boson mass $(m_Z )$ has also been excluded by LEP experiments from Z-boson decay upto $|V_{\alpha i} |^2 \,\approx\, 10^{-5}$ \cite{Adriani:1992pq,Abreu:1996pa,L3}. These constraints are taken care of in our analysis by keeping the mass of the lightest heavy neutrino to be greater than or equal to 200 GeV.

\subsection{Bounds on the Scalar Sector}
\item[$\bullet$ Constraints on scalar potential couplings from perturbativity and unitarity] 


For the radiatively improved Lagrangian of our model to be perturbative, we should have \cite{Lee:1977eg,Cynolter:2004cq},
\be \lambda (\Lambda)\, 
 <\, \frac{4\pi}{3} \,\,\, ; \,\,\, |\kappa(\Lambda)|\, < \,8\pi \,\,\,;\,\,\, |\lambda_{S}(\Lambda)|\, < \, 8\pi\ee at all scales and the values of the couplings at any scale $\Lambda$ are evaluated using the RG equations. The parameters of the scalar potential (see eqn. (\ref{eqpot1})) of this model are also constrained by the unitarity of the scattering matrix (S-matrix). At very high field values, one can obtain the scattering matrix by using various scalar-scalar, gauge boson-gauge boson, and scalar-gauge boson scattering amplitudes. Using the equivalence theorem~\cite{equivalence1,equivalence2,equivalence3}, we have reproduced the scattering matrix (S-matrix) for this model ~\cite{Cynolter:2004cq}. The unitarity demands that the eigenvalues of the S-matrix should be less than $8\pi$. The unitary bounds are given by, 
\bea
\lambda \leq 8 \pi  \quad {\rm and} \quad \Big| 12 {\lambda}+{\lambda_S} \pm \sqrt{16 \kappa^2+(-12 {\lambda}+{\lambda_S})^2}\Big| \leq 32 \pi.\nonumber
\eea

\item[$\bullet$ Dark matter constraints]

The parameter space for the scalar sector should also satisfy the Planck and WMAP imposed dark matter relic density constraint \cite{Ade:2015xua},
  \be \Omega_{DM}h^2  \, = \,  0.1198 \, \pm \, 0.0026. \ee

In addition, the invisible Higgs decay width and the recent direct detection experiments, in particular, the LUX-2016~\cite{Akerib:2016vxi} data and the indirect Fermi-LAT data\cite{TheFermi-LAT:2015kwa}  restrict the arbitrary Higgs portal coupling and the dark matter mass~\cite{Khan:2014kba, Athron:2017kgt}. 

 Since the extra fermions are heavy ($\gtrsim$ 200 GeV), for low dark matter mass (around 60 GeV), the dominant (more than 75 $\%$) contributions to the relic density is from the $SS\rightarrow b \bar{b}$ channel. The channels $SS\rightarrow V,V^*$ also  contribute to the relic density where $V$ stands for the vector bosons $W$ and $Z$, $V^*$ indicates the virtual particle which can decay into the SM fermions. In this mass region, the value of the Higgs portal coupling $\kappa$ is $\mathcal{O}$($10^{-2}$) to get the relic density in the right ballpark and simultaneously satisfying the other experimental bounds. However, this region is not of much interest to us since such a small coupling will not contribute much to the running of $\lambda$ and hence will not affect the stability of the EW vacuum much. The LUX-2016 data~\cite{Akerib:2016vxi} has ruled out the dark matter mass region $\sim$ $70-500$ GeV.
    
    If we consider  $M_{DM} >> M_t$, the annihilation cross-section is proportional to $\frac{\kappa^2}{M_{DM}^2}$, which ensures that the relic density band in $\kappa-M_{DM}$~\cite{Khan:2014kba} plane is a straight line. In this region, one can get the right relic density if the ratio of dark matter mass to the Higgs portal coupling $\kappa$ is $\sim$ 3300. In this case, the dominant contributions to the dark matter annihilation channel are $SS\rightarrow hh, t \bar{t}$, VV.
    
    We use FeynRules \cite{Alloul:2013bka} along with micrOMEGAs \cite{Belanger:2010gh,Belanger:2013oya} to compute the relic density of the scalar DM. We have checked that the contribution of annihilation into extra fermions is very small. However this could be significant for dark matter mass $\gtrsim$ 2.5 TeV provided the Yukawa couplings are large enough. But, in the stability analysis discussed in section~\ref{TunnelingPhase}, we will see that the dark matter mass $\gtrsim$ 2.5 TeV requires the value of $\kappa \gtrsim 0.65 $ which  violates the perturbativity bounds before the Planck scale. Thus, we consider the dark matter mass in the range $\sim$ 500 GeV - 2.5 TeV with $\kappa$ in the range $\sim$ 0.15 to 0.65. It is to be noted that in the presence of the singlet fermions the value of $\kappa$($M_Z$) and hence $M_{DM}$ for which the perturbativity is not obeyed will also depend upon the value of Tr $[Y_\nu^\dag Y_\nu]$. This will be discussed in the next section.

\end{description}

\section{Results}
In this section, we present our results of the stability analysis of the electroweak vacuum in the two seesaw scenarios.  We confine ourselves to the normal hierarchy. The results for the inverted hierarchy are not expected to be very different \cite{Khan:2012zw}. We have used the  package SARAH\cite{Staub:2013tta} to do the RG analysis in our work.
\subsection{Inverse Seesaw Model}
For the inverse seesaw model, the input parameters are the entries of the matrices $Y_\nu$, $M_S$ and $M_{\mu}$.
Here $Y_\nu$ is a complex $3 \times 3$ matrix. $M_S$ is a real $3 \times 3$ matrix and $M_{\mu} $ is a $3 \times 3$ diagonal matrix with real entries.
We vary the entries of various mass matrices  in the range $10^{-2} \, < \, M_\mu \, < \, 1$ keV and $0 \, < M_R \, < \, 5\times10^4 $ GeV.
This implies a a heavy neutrino mass of maximum upto a few TeV. 
With these input parameters, we search for parameter sets  consistent with the low energy data using the downhill simplex method \cite{Press:1996}.  
We present in table \ref{benchmark}, some representative outputs consistent with data for three benchmark points. In this table $Tr[Y_{\nu}Y^{\dag}_{\nu}]$ is an input. As a consistency check, we also give the value of  $Br(\mu \rightarrow e~\gamma)$.

\begin{table}[ht]
 $$
 \begin{array}{|c|c|c|c|}
 \hline {\mbox {Parameter} }& BM-I & BM-II& BM-III\\
 \hline
  \Delta m^2_{21}/10^{-5} eV^{2}&            8.0891 &  7.8228 & 7.6277\\
   \Delta m^2_{31}/10^{-3} eV^{2}&             2.4391 & 2.5046  & 2.4078\\
          \sin^2\theta^L_{12}&            0.2710  &    0.3429   & 0.3449\\
          \sin^2\theta^L_{23}&            0.3850  &  0.3850   &   0.4102\\
          \sin^2\theta^L_{13}&            0.0239  &  0.0229  & 0.0238\\
                \delta_{PMNS}&            1.1173   &  1.4273  &  1.1715\\
               \phi_1,\phi_2 &            2.5187,            2.9377 &  2.9384,            3.1379   & 0.4264,            0.7426\\
        m_i/10^{-1} ~eV&              0.10,              0.13,             0.511&   0.23,              0.25,             0.558  &   0.10,              0.13,             0.507\\
              M_i~GeV&        200.77,            200.77,           461.159,  ,& 210.01,            210.01,           487.284,&   200.00,            200.00,           332.993
               \\&         461.16,           1744.67,          1744.669 & 487.28,           1451.34,          1451.344 & 332.99,           3568.87,          3568.869\\
    Tr[Y_{\nu}Y^{\dag}_{\nu}]&            0.1&0.2&0.3\\
Br(\mu \rightarrow e~\gamma)&          0.731 \times 10^{-16}  & 0.1 \times 10^{-16}  &   0.13 \times 10^{-15}\\
 \hline
 \end{array}
 $$
 \label{table f}\caption{\small{ Output values for three different benchmark points for inverse seesaw model satisfying all the low energy
 constraints
 }}\label{benchmark}\end{table}

\subsubsection{Vacuum Stability}
\begin{figure}[h]
    \centering
    \begin{subfigure}[b]{0.45\textwidth}
        \includegraphics[width=\textwidth]{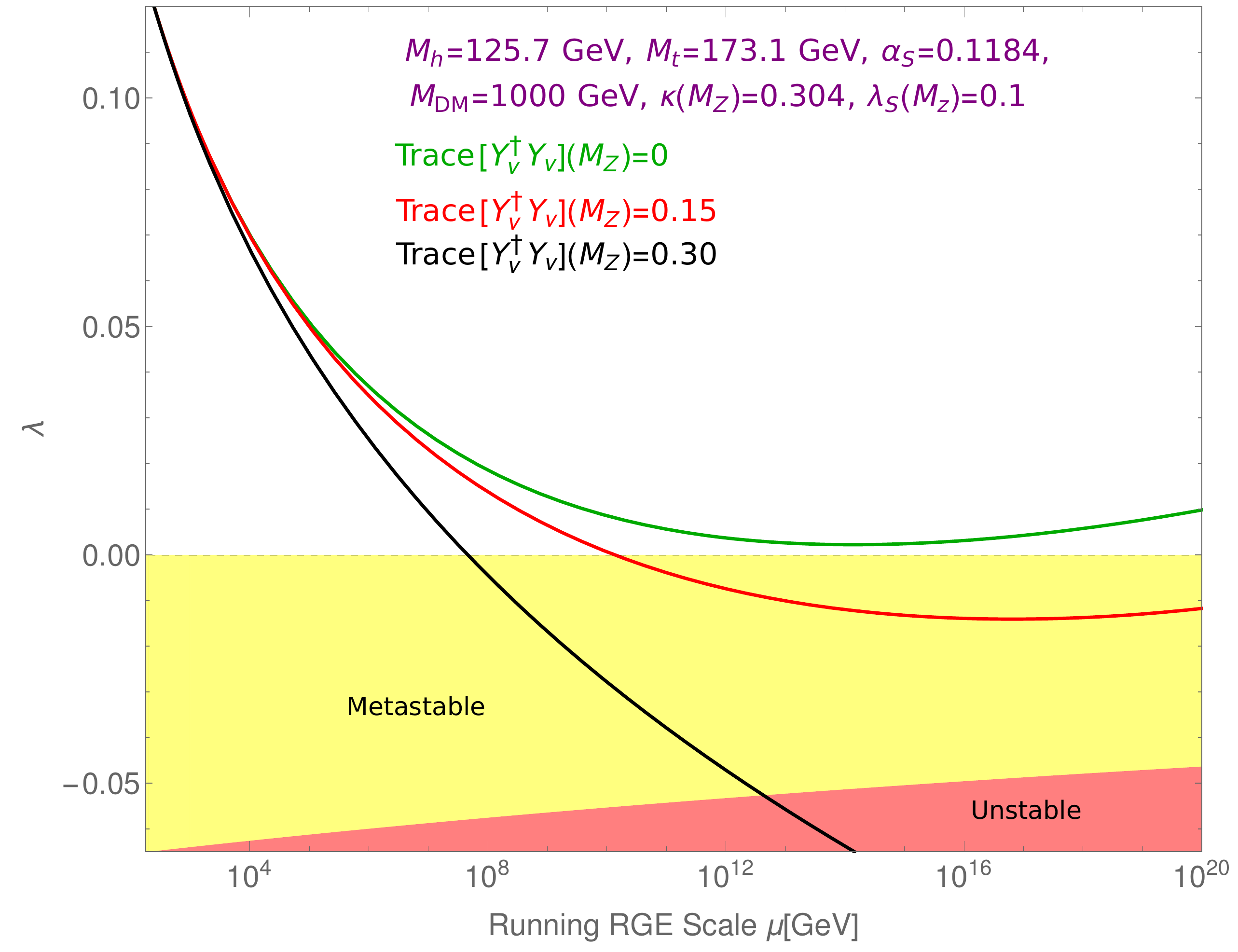}
        \caption{ \centering Running of $\lambda$  for different values\\of Tr $[Y_\nu^\dag Y_\nu]$ and a fixed value of $\kappa$} \label{1a}
    \end{subfigure}
    ~ 
    \begin{subfigure}[b]{0.45\textwidth}
        \includegraphics[width=\textwidth]{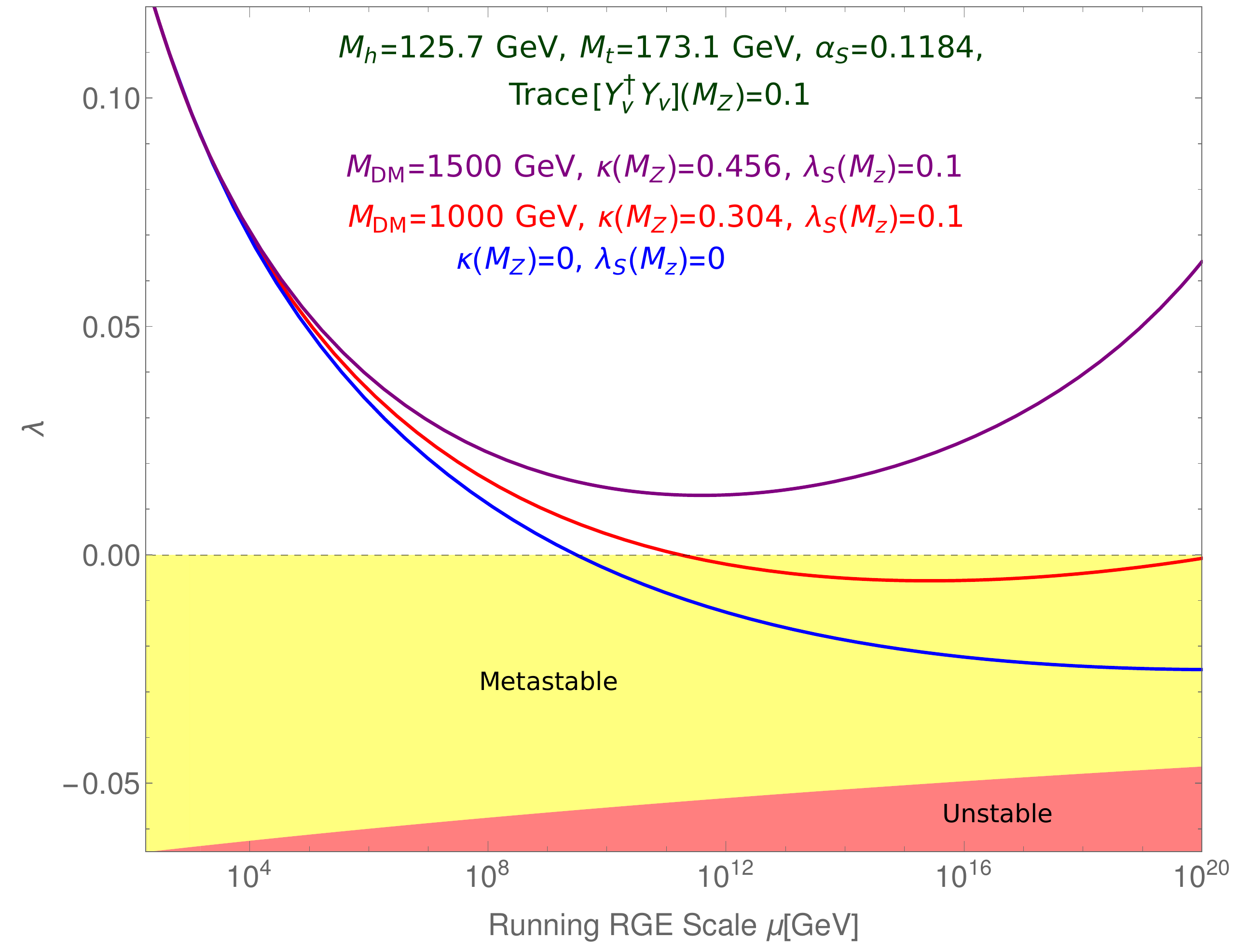}
        \caption{ \centering Running of $\lambda$ for a fixed  value\\of Tr $[Y_\nu^\dag Y_\nu]$ and different values of $\kappa$} \label{1b}
            \end{subfigure}
    
       \begin{subfigure}[b]{0.45\textwidth}
        \includegraphics[width=\textwidth]{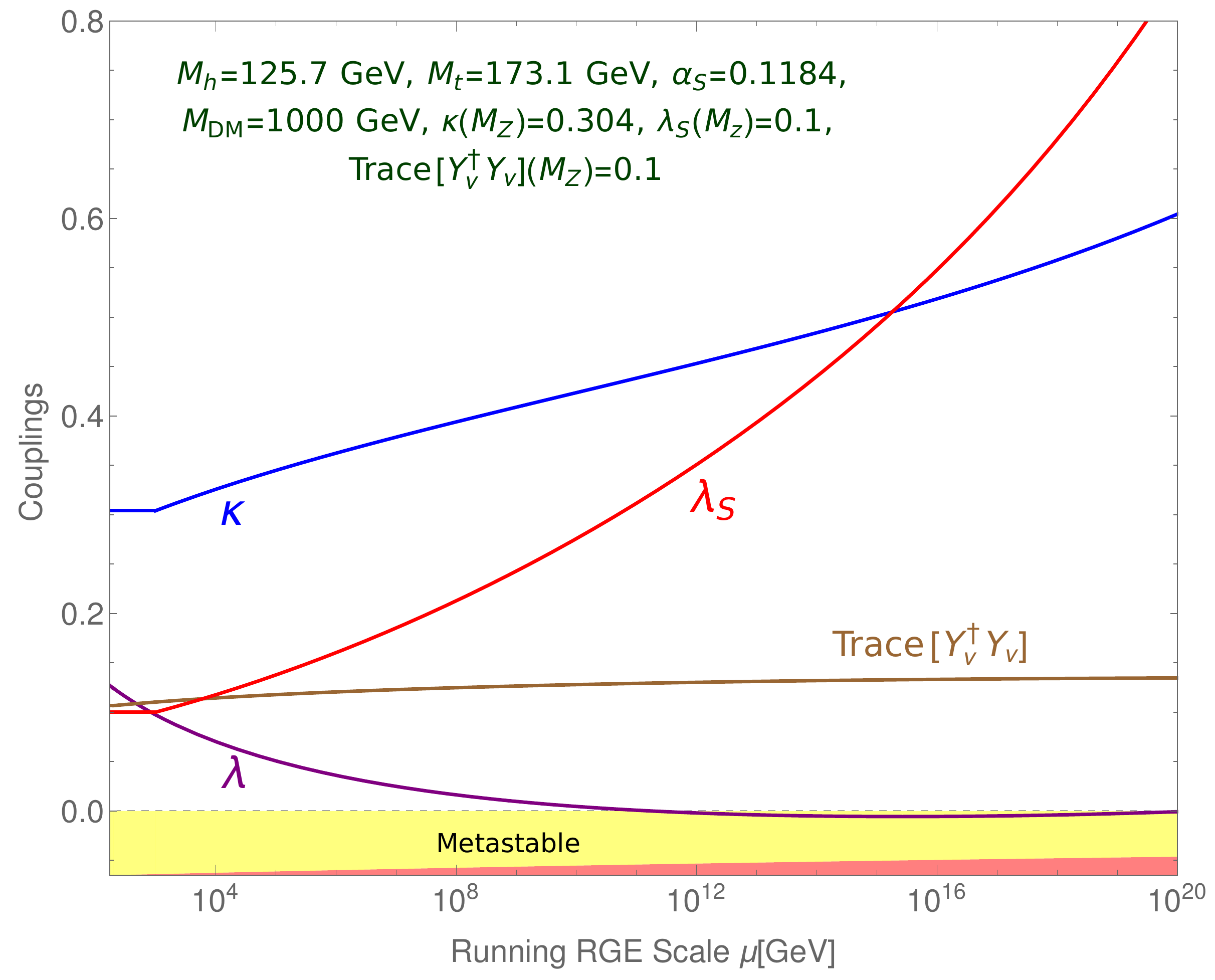}
        \caption{\centering  Running of the couplings with energy for dark matter mass of 1000 GeV} \label{1c}
    \end{subfigure}
    ~ 
    \begin{subfigure}[b]{0.45\textwidth}
        \includegraphics[width=\textwidth]{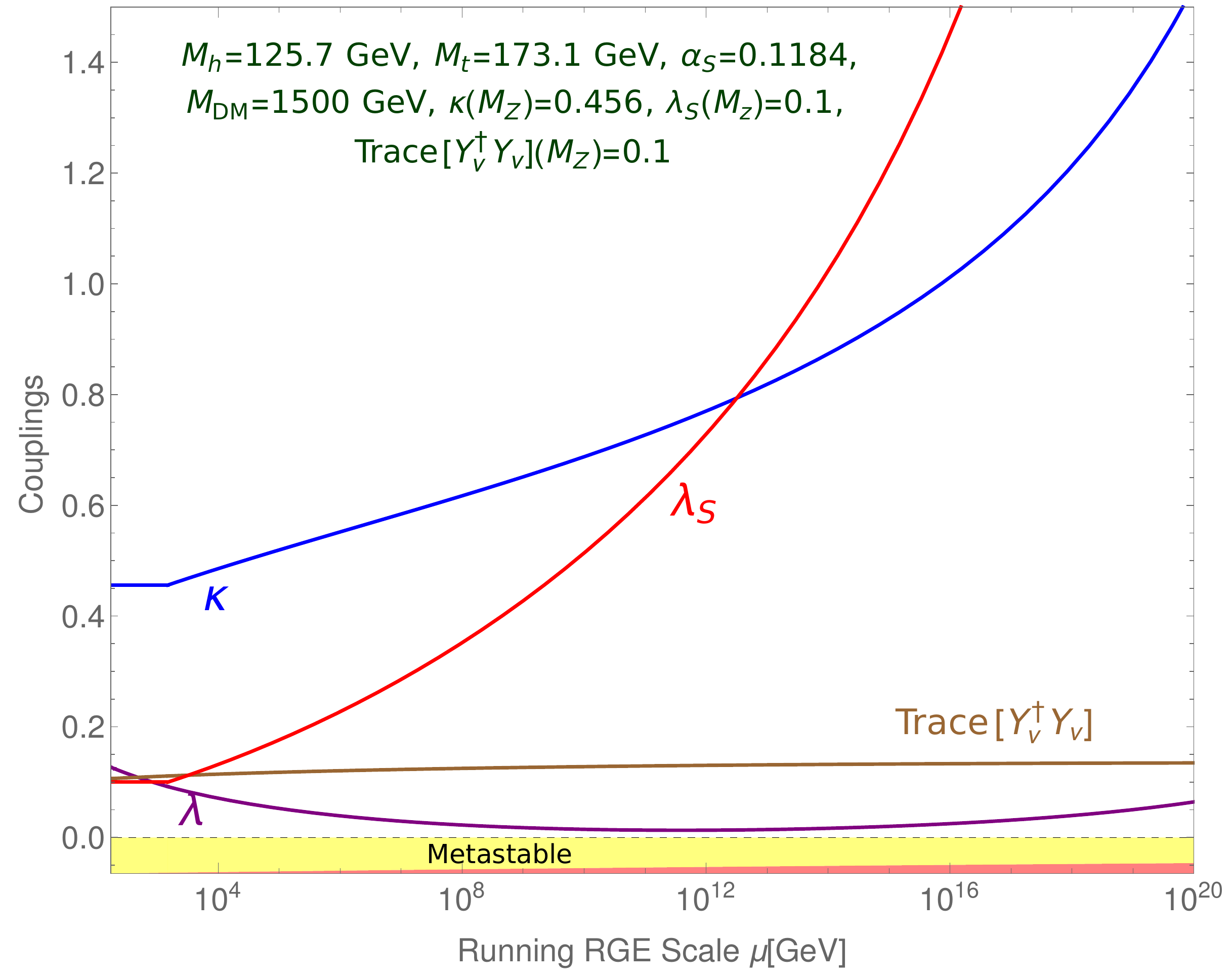}
        \caption{\centering  Running of the couplings with energy for dark matter mass of 1500 GeV} \label{1d}
            \end{subfigure}
    
    \caption{\centering  Running of the couplings with the energy scale in the Inverse seesaw model }\label{1}
\end{figure}

In fig.(\ref{1}), we display the running of the couplings for various benchmark points in the ISM. 
In fig.(\ref{1a}), we have shown the variation in the running of the Higgs quartic coupling $\lambda$ for different values of Tr $[Y_\nu^\dag Y_\nu]$ (0, 0.15 and 0.30) for a
fixed value of the Higgs portal coupling $\kappa$ = 0.304. We have chosen the DM mass $M_{DM}$=1000 GeV to get the relic density in the right ballpark. As $\lambda_S$ doesn't alter the relic density, we have fixed it's value at $0.1$ for all the plots in this paper.
We can see that for Tr $[Y_\nu^\dag Y_\nu] = 0$ , i.e., without the right handed neutrinos, the EW vacuum remains absolutely stable upto the Planck scale (green line)  and for the large values of Tr $[Y_\nu^\dag Y_\nu]$, the  EW vacuum goes towards the instability (Higgs quartic coupling becomes negative around $\Lambda_I ~\sim~10^{10}$ GeV (red line) and $\Lambda_I ~\sim~10^{8}$ GeV (black line)) region. 
 
 In fig.(\ref{1b}), we plot the running of $\lambda$ for a fixed value of Tr $[Y_\nu^\dag Y_\nu]$ = 0.1 and different sets of $k$ and $~ M_{DM}$. 
It is seen that for a larger value of $\kappa = 0.45$ with  $M_{DM}$ = 1500 GeV, the EW vacuum remains stable upto Planck scale (purple line).  For $\kappa = 0.304$ with  $M_{DM}$ = 1000 GeV, the quartic coupling $\lambda$ (red line) becomes negative around  $\Lambda_I ~\sim~10^{11}$ GeV  and in the absence of the singlet scalar field, i.e., for $\kappa = 0 $, $\lambda_{S}=0 $ (blue line), $\lambda$ becomes negative  around  $\Lambda_I ~\sim~10^{9}$ GeV and the vacuum goes to the metastability region.

 In figs.(\ref{1c}) and (\ref{1d}), we have shown the running of all the three scalar quartic couplings, $\lambda,\, \kappa$ and $\lambda_S$ and Tr$[Y_\nu^\dag Y_\nu]$ for $ (M_{DM}, \,\kappa)$  = (1000 GeV, 0.304)  and (1500 GeV, 0.456) respectively. It can be seen that the values of $\lambda_s$ and $\kappa$ increases considerably with the energy scale and can reach the perturbativity bound at the Planck scale depending upon the initial values of $\kappa$ and $\lambda_S$ at $M_Z$. Here for $\lambda_S$=0.1, the maximum allowed value of $\kappa$ will be 0.58 from the perturbativity.
 The value of Tr$[Y_\nu^\dag Y_\nu]$ increases only slightly with the energy scale and the value of $\lambda_S$ increases faster for larger value of $\kappa$.

\subsubsection{Tunneling Probability and Phase Diagrams}
\label{TunnelingPhase}

 The present central values of the SM parameters, especially the top Yukawa coupling $y_t$ and strong coupling constant $\alpha_s$ with Higgs mass $M_{h} \,\approx\, 125.7$ GeV suggest that the 
 beta function of the Higgs quartic coupling $\beta_{\lambda} (\equiv dV(h)/dh)$ goes from negative to positive around $10^{15}$ GeV \cite{Alekhin:2012py,Buttazzo:2013uya}.
 This implies that there is an extra deeper minima  situated at that scale. So there is a finite probability
that the electroweak vacuum might tunnel into that true (deeper) vacuum. But this tunneling probability is not large enough and hence the life time 
of the EW vacuum remains larger than the age of the universe. This implies that the EW vacuum is metastable in the SM. The expression for the tunneling
probability at zero temperature is given by \cite{Isidori:2001bm,Espinosa:2007qp},
\be {\mathcal P}_0 \, =\,  V_U\,\Lambda_B^4\, \textrm{exp} \, \Big( -\frac{8\pi^2}{3\,|\lambda(\Lambda_B)|}  \Big)   \ee
 where $\Lambda_B $ is the energy scale at which the action of the Higgs potential is minimum. $V_U$ is the volume of the past light cone taken as $\tau_U^4$, where $\tau_U$ is
 the age of the universe ($\tau_U \,=\,4.35\,\times \,10^{17} $ sec)\cite{Ade:2015xua}.  In this work we have neglected the loop corrections and gravitational correction to the action of the Higgs potential~\cite{Isidori:2007vm}. For the vacuum to be metastable, we should have $ {\mathcal P}_0 < 1$ which implies that \cite{Khan:2014kba},
  \be 0 \,> \, \lambda(\mu) \,> \, \lambda_{min}(\Lambda_{B}) \, = \, \frac{-0.06488}{1-0.00986\,\textrm{ln}\,(v/\Lambda_{B})},  \label{lambdamin} \ee 
whereas the situation $\lambda(\mu) \,< \, \lambda_{min}(\Lambda_{B})$ leads to the unstable EW vacuum. In these regions, $\kappa$ and $\lambda_S$ should always be positive  to get the scalar potential bounded from below ~\cite{Khan:2014kba}.
In our model, the EW vacuum shifts towards stability/instability depending upon the  new physics parameter space for the central values of 
$M_h=125.7$ GeV, $M_{t}=173.1$ GeV and $\alpha_s=0.1184$ and there might be an extra minima around $10^{12-17}$ GeV.

\begin{figure}

    \includegraphics[scale=0.4]{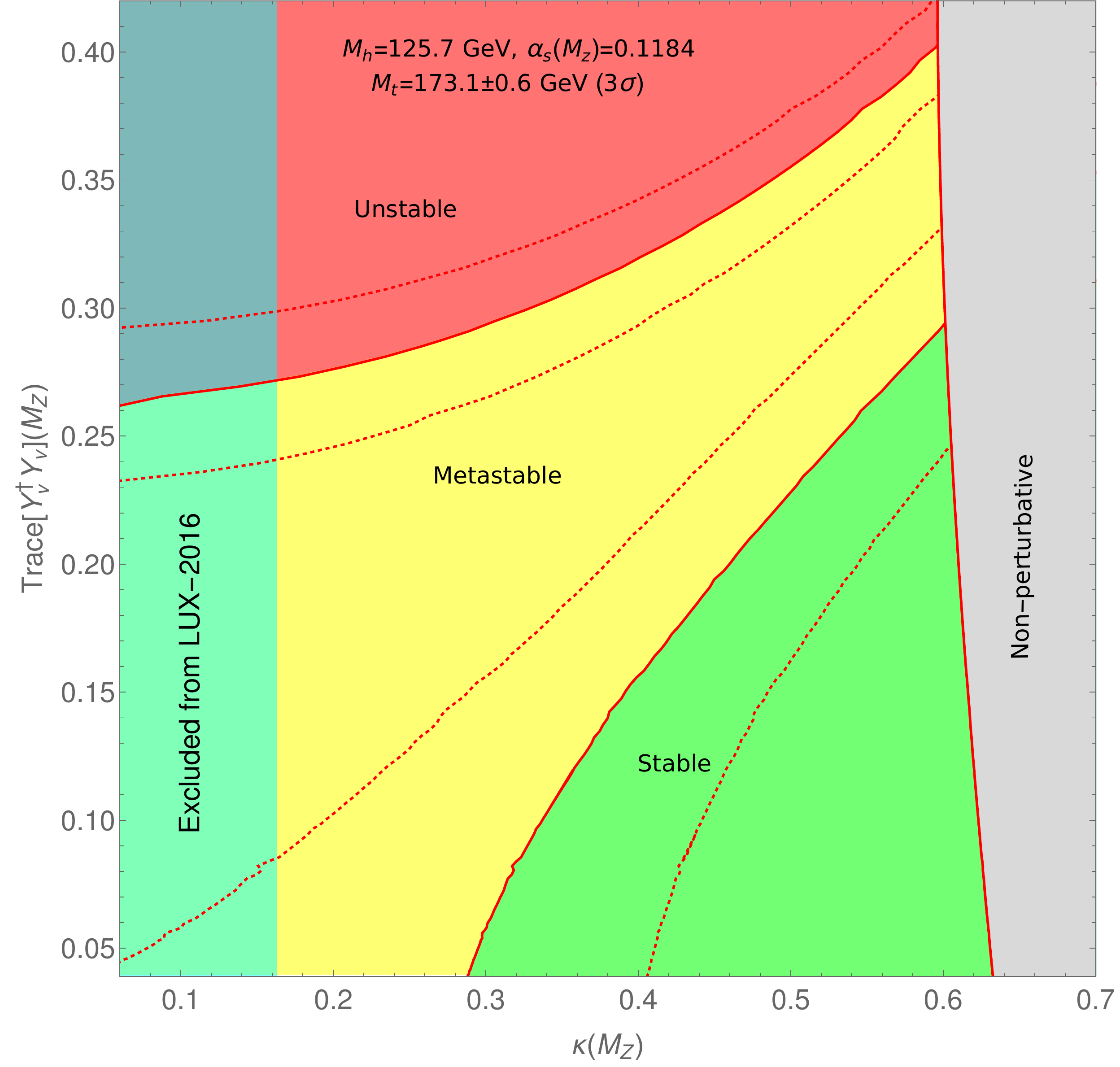}
    \caption{Phase diagram in the Tr$[Y_\nu^\dag Y_\nu]$ - $\kappa$ plane. We have fixed all the entries 
    of $Y_\nu$ except for $(Y_\nu)_{33}$. The three boundary lines (two dotted and a solid) 
    correspond to $M_t \, = \, 173.1 \,\pm \, 0.6$ GeV (3$\sigma$) and we have taken $\lambda_S (M_Z) = 0.1$.} \label{PDISM}
\end{figure}

In fig.(\ref{PDISM}), we have given the phase diagram in the $ \textrm{Tr}\, [Y_\nu^\dag Y_\nu] \,-\,\kappa$ plane. 
The line separating the stable region and the metastable region is obtained when the two vacuua are at the same depth, 
i.e., $\lambda(\mu) \, = \, \beta_\lambda(\mu) \, = \, 0$. The unstable and the metastable regions are separated by the
boundary line where $\beta_\lambda(\mu) \, = \, 0$ along with $\lambda(\mu) \, = \, \lambda_{min}(\Lambda_B)$, as defined in eqn. (\ref{lambdamin}).  For simplicity, we have plotted fig.(\ref{PDISM}) (also fig.(\ref{1})) by fixing all the eight entries of the $3 \times 3$ complex matrix $Y_\nu$,
but varying only the $(Y_{\nu})_{ 33}$ element to get a smooth phase diagram.
From fig. (\ref{PDISM}), it could be seen that the values of $\kappa$ beyond $\sim$ 0.58 are disallowed by 
perturbativity bounds and those below $\sim \, 0.16$ are disallowed by the direct detection bounds from LUX-2016~\cite{Akerib:2016vxi}. 
Note that the vacuum stability analysis of the inverse seesaw model done in reference \cite{Rose:2015fua} had found that 
the parameter space with $\textrm{Tr}\,[ Y_\nu^\dag Y_\nu] \, > \, 0.4 $ were excluded by vacuum metastability constraints.
Whereas, in our case, fig.(\ref{PDISM}) shows that the parameter space with $\textrm{Tr}\,[ Y_\nu^\dag Y_\nu] \, \gtrsim \,  0.25 $ are 
excluded for the case when there is no extra scalar.  The possible reasons could be that we have kept the maximum value of the heavy neutrino mass to be around a few TeV, whereas the authors of \cite{Rose:2015fua} had considered heavy neutrinos 
as heavy as 100 TeV. Obviously, considering larger thresholds would allow us to consider large value of Tr$[Y_\nu^\dag Y_\nu]$ as the corresponding couplings
will enter into RG running only at a higher scale. Another difference with the analysis of \cite{Rose:2015fua} is that we have fixed 8 of the 9 entries of the Yukawa coupling matrix $Y_\nu$. Also, varying all the 9 Yukawa couplings will give us more freedom 
and the result is expected to change. The main result that we deduce from this plot is the effect of $\kappa$ on the maximum allowed value of
Tr $[Y_\nu^\dag Y_\nu]$, which increases from 0.26 to 0.4
for a value of $\kappa$ as large as 0.6. In addition, we see that the upper bound on $\kappa(M_Z)$ from perturbativity at the Planck scale decreases from
0.64 to 0.58 as the value of Tr[$Y_\nu^\dag Y_\nu$] changes from 0 to 0.44. This can be explained from the expression of the $\beta_{\kappa}$ in eqn. (\ref{anom}) which shows that  $[Y_\nu^\dag Y_\nu]$ affect the
 running $\kappa$ positively through the quantity $T$. Since $M_{DM}$ $\sim$ 3300 $\kappa$ for $M_{DM}>> M_{t}$,  
the mass of dark matter for which perturbativity is valid, decreases with increase in the value of the Yukawa coupling.
\begin{figure}[h]
    \begin{subfigure}[b]{0.465\textwidth}
        \includegraphics[width=\textwidth]{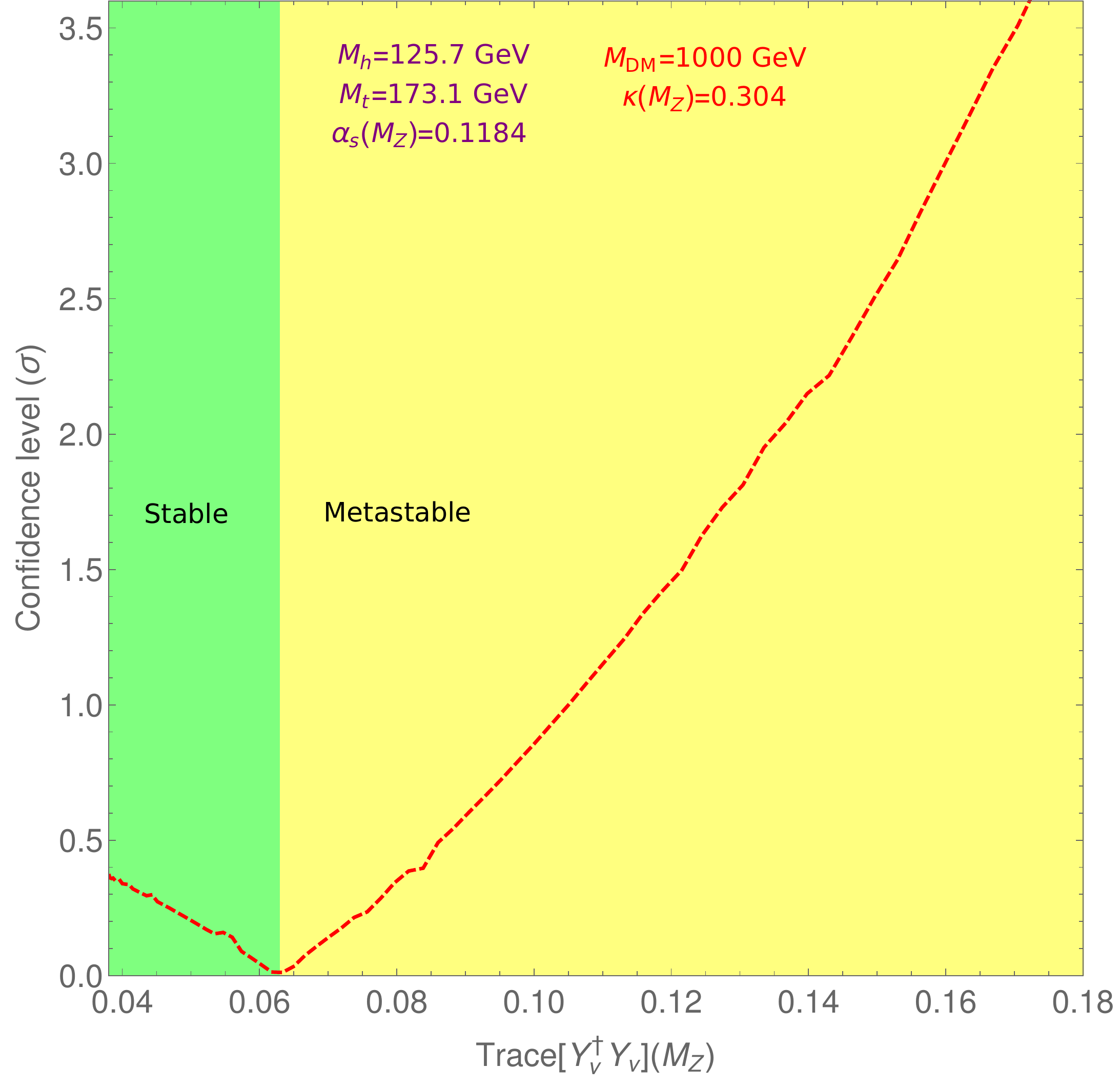}
        \caption{ \centering  ($\kappa, \, M_{DM}$) = (0.304, 1000 GeV)    }\label{3a}
         \end{subfigure}
       \begin{subfigure}[b]{0.45\textwidth}
        \includegraphics[width=\textwidth]{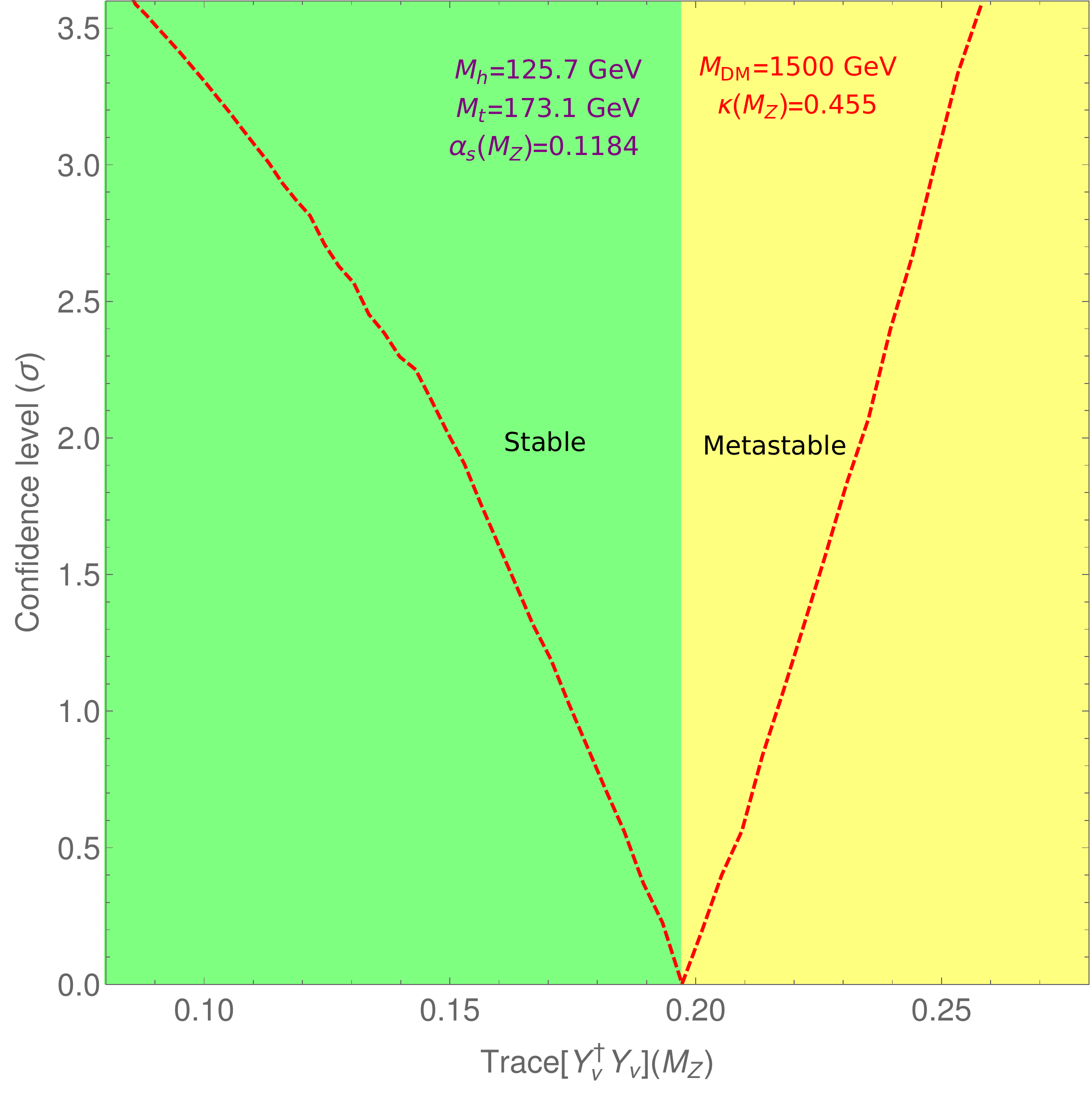}
        \caption{\centering  ($\kappa, \, M_{DM}$) = (0.455, 1500 GeV) }\label{3b}
                    \end{subfigure}
  \caption{Dependence of confidence level at which the EW vacuum stability is excluded/allowed on Tr$[Y_\nu^\dag Y_\nu]$ for 
        two different values of $\kappa$ and $M_{DM}$. We have taken $\lambda_S(M_Z) = 0.1$}\label{CDISM}
\end{figure}

\subsubsection{Confidence level of vacuum stability}
As we have seen that the stability of the electroweak vacuum changes due to the presence of new physics  and hence it becomes important to demonstrate the change in the confidence level at which stability is excluded or allowed (one-sided) \cite{Khan:2014kba, Khan:2015ipa, Khan:2016sxm}. 
In particular, it will provide a quantitative measurement of (meta)stability in the presence of new physics. In fig.(\ref{CDISM}), we graphically show how the confidence level at which stability of electroweak
vacuum is allowed/excluded depends on new Yukawa couplings of the heavy fermions for the inverse seesaw model
in the presence of the extra scalar (dark matter) field. 
We have plotted the dependence of confidence level against the trace of the Yukawa coupling,
Tr$[Y_\nu^\dag Y_\nu]$ for fixed values of Higgs portal coupling $\kappa\,=\,0.304$ in fig.(\ref{3a}).
Here, the dark matter mass $M_{DM}=1000$ GeV is dictated by $\kappa$ to obtain the correct relic density.
Similar plot with a higher value of $\kappa=0.455$ with dark matter mass $M_{DM}=1500$ GeV is shown in fig.(\ref{3b}).
In this case the electroweak vacuum is absolutely stable for a larger parameter space. For a particular set of values of the model parameters
$M_h \, = \, 125.7$ GeV, $M_t \, = \, 173.1$ GeV, $\alpha_s (M_z) \, = \, 0.1184$ and $\kappa$,
 the confidence level (one-sided) at which the electroweak  vacuum
is absolutely stable (green region) decreases with the increase of Tr$[Y_\nu^\dag Y_\nu]$ and becomes zero for Tr$[Y_\nu^\dag Y_\nu]=0.06$ in
fig.(\ref{3a}) and Tr$[Y_\nu^\dag Y_\nu]=0.20$ in fig.(\ref{3b}). The confidence level at which the absolute stability of
electroweak vacuum is excluded (one-sided)
increases with the trace of the Yukawa coupling in the yellow region. 
 
\subsection{Minimal Linear Seesaw Model}\label{MLSMsub}
In the minimal linear seesaw case, the Yukawa coupling matrices $Y_\nu$ and $Y_s$ can be completely determined in terms of the oscillation parameters apart from the overall coupling constant $y_\nu$ and $y_s$ respectively \cite{Gavela:2009cd}.  
For normal hierarchy, in MLSM, the Yukawa coupling matrices $Y_\nu$ and $Y_S$ can be parametrized as,
 \be \label{Ynu1}  Y_\nu \, =\, \frac{y_\nu}{\sqrt{2}} \,\Big( \sqrt{1+\rho}\,U_3^\dag \,+\, \textrm{e}^{i\frac{\pi}{2}}\sqrt{1-\rho}\,U_2^\dag \Big) \ee
 
  \be   \label{Ys1} Y_s \, =\, \frac{y_s}{\sqrt{2}} \,\Big( \sqrt{1+\rho}\,U_3^\dag \,+\, \textrm{e}^{i\frac{\pi}{2}}\sqrt{1-\rho}\,U_2^\dag \Big) \ee
 
 where \be \rho\,=\,\frac{\sqrt{1+r}-\sqrt{r}}{\sqrt{1+r}+\sqrt{r}} .\ee
 
Here, $U_i$'s are the columns of the unitary PMNS matrix $U_\nu$ and $r$ is the ratio of the solar 
and the atmospheric mass squared differences. This parametrization makes the vacuum stability analysis in the minimal linear seesaw model much more easier since there are only two independent parameters $y_{\nu}$ and $M_{N}$ in the fermion sector, where $M_N$ is the degenerate mass of the two heavy neutrinos (the value of $y_s$ being very small $ \mathcal{O}(10^{-11})$).  A detailed analysis has already been performed in reference  \cite{Khan:2012zw}. 
Here, we are interested in the interplay between the $Z_2$ odd singlet scalar and singlet fermions in the vacuum stability analysis.

\begin{figure}[h]
    \centering
    \begin{subfigure}[b]{0.45\textwidth}
        \includegraphics[width=\textwidth]{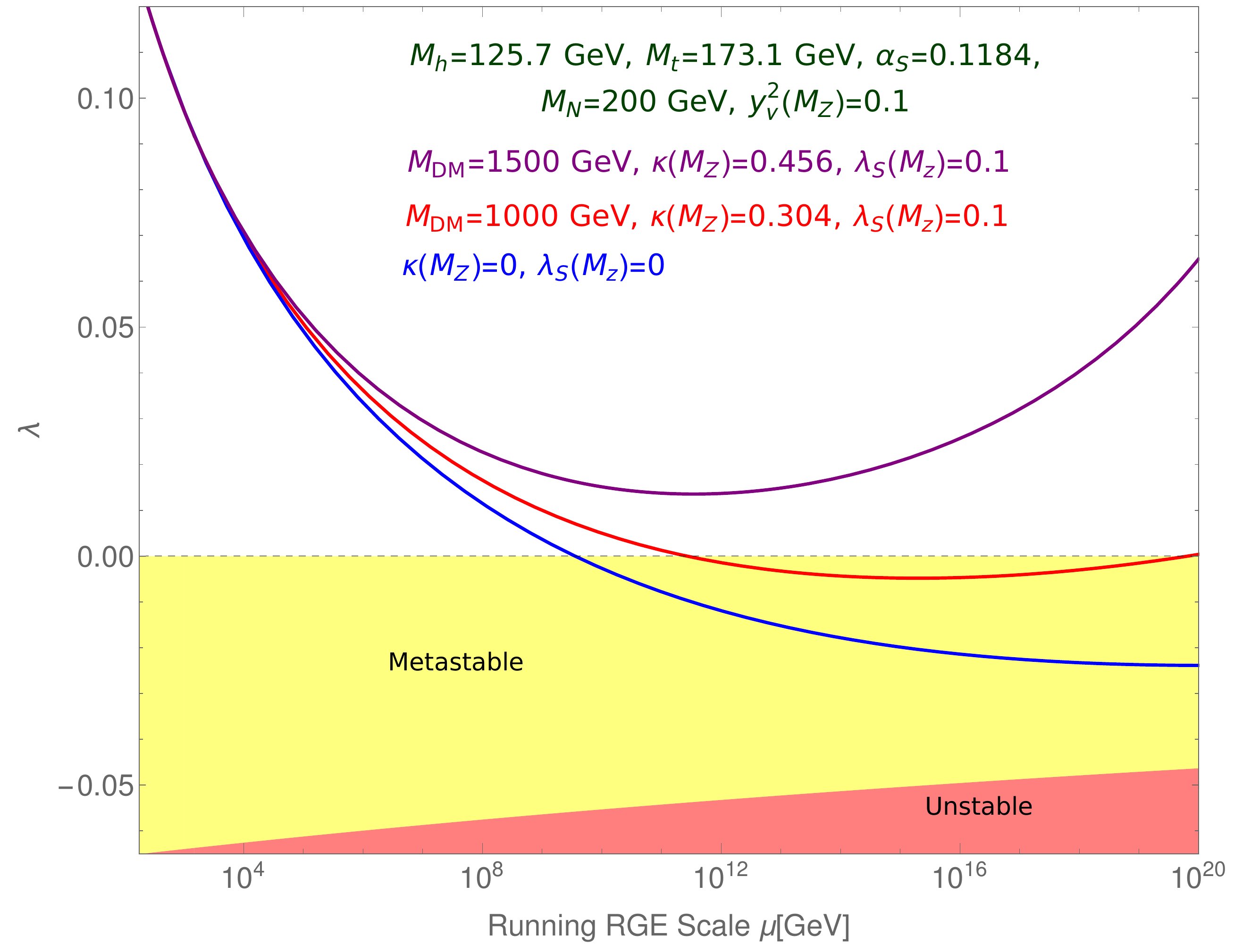}
        \caption{\centering }\label{4a}
         \end{subfigure}
       \begin{subfigure}[b]{0.45\textwidth}
        \includegraphics[width=\textwidth]{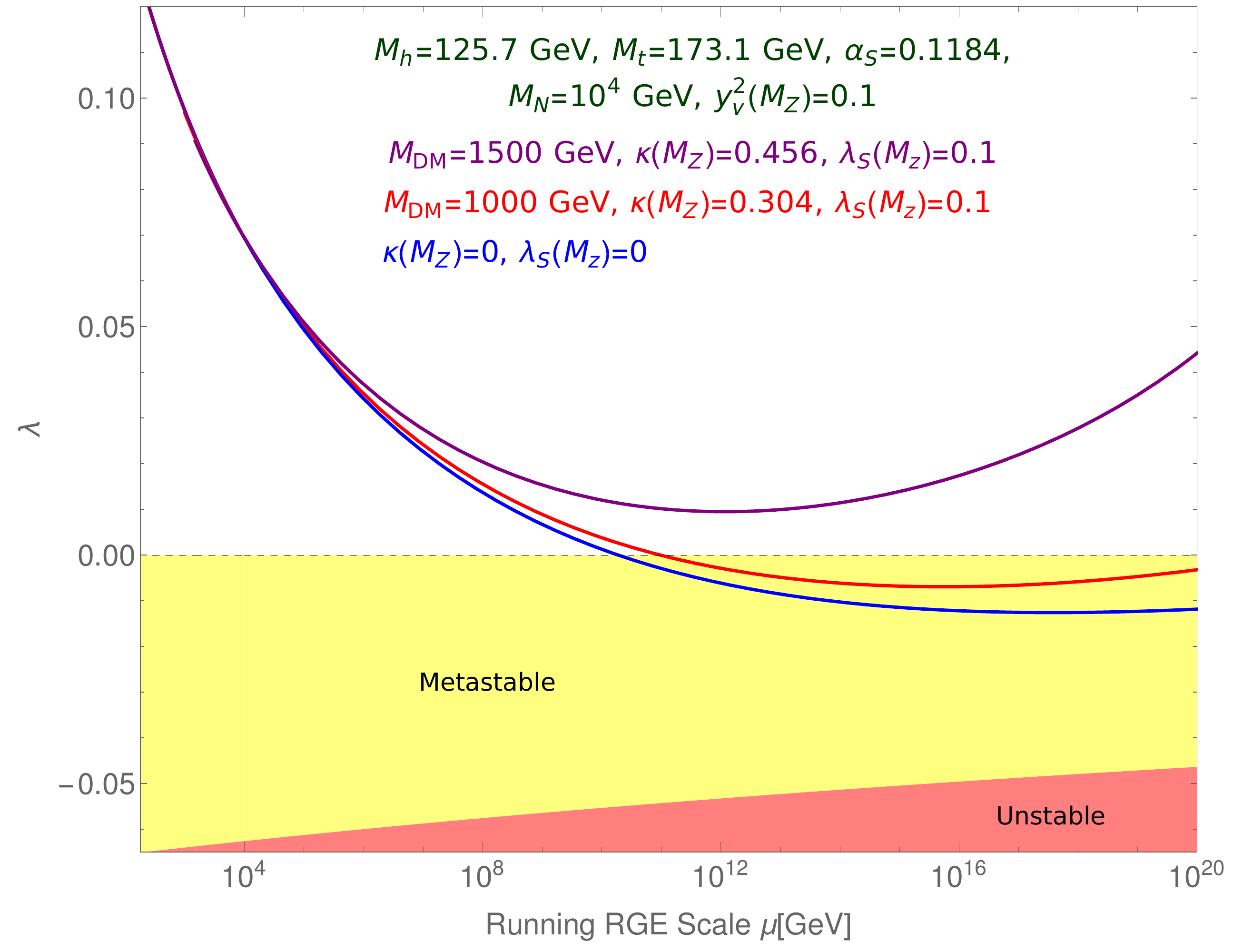}
        \caption{\centering }\label{4b}
                    \end{subfigure}
 
 \begin{subfigure}[b]{0.45\textwidth}
        \includegraphics[width=\textwidth]{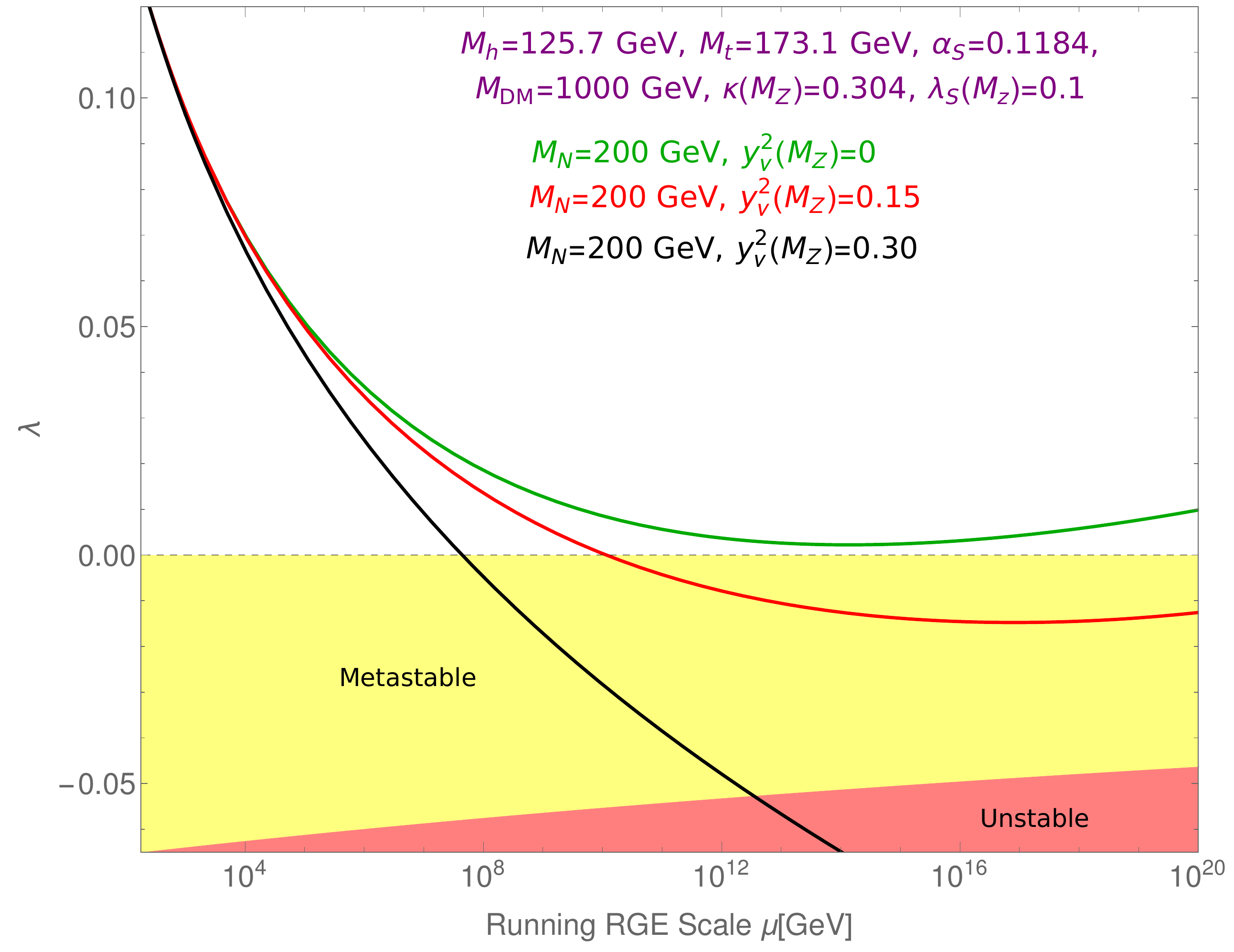}
        \caption{\centering }\label{4c}
         \end{subfigure}
       \begin{subfigure}[b]{0.45\textwidth}
        \includegraphics[width=\textwidth]{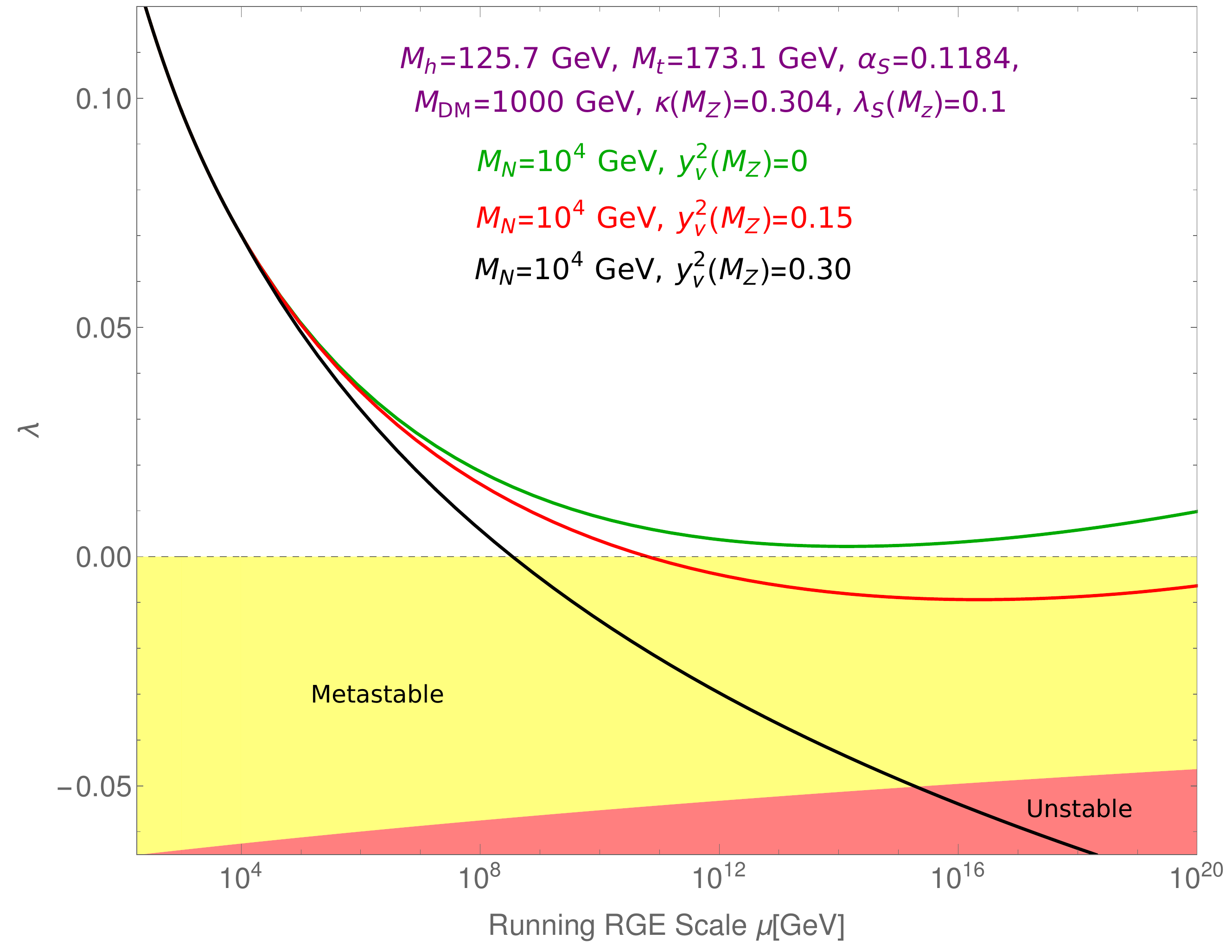}
        \caption{\centering }\label{4d}
        \end{subfigure}
 
        \caption{Running of the quartic coupling $\lambda$ in MLSM with extra scalar for two different values of $M_N$. 
        In the upper panel, the three lines are for different  values of $M_{DM}$ and $\kappa$ whereas in the lower panel,
        they are for different  values of $y_\nu$ and fixed values of $M_{DM}$ and $\kappa$.}
\label{linearcouprun}\end{figure}

In fig.(\ref{linearcouprun}), we have plotted the running of the Higgs quartic coupling $\lambda$ with the energy scale 
$\mu$ upto the Planck scale. The figs.(\ref{4a}) and (\ref{4b})  show the running of $\lambda$ for different values of $k$ (0.0, 0.304, 0.456) and $M_{DM}$ (0,1000 GeV, 1500 GeV), for $M_N$=200 GeV and $M_N$ = $10^4$ GeV respectively for a fixed value of $y_{\nu}^2$=0.1. Comparing these two plots, we can see that $\lambda$ tends to go to the instability region faster  for smaller values of the heavy neutrino mass. So, the EW vacuum is more stable for larger values of $M_{N}$, because the effect of extra singlet fermion in the running of $\lambda$ enters at a higher value. We also find that as the value of $\kappa$ increases from 0 to 0.304, the electroweak vacuum becomes metastable at a higher value of the energy scale. For $\kappa$=0.456 the electroweak vacuum becomes stable upto the Planck scale even in the presence of the singlet fermions. 

 Figs.(\ref{4c}) and (\ref{4d}) display the running of  $\lambda$ for different values of $y_{\nu}^2$ (0.0, 0.15, 0.3) and for fixed values of $k$=0.304 and $M_{DM}$=1000 GeV, for $M_N$= 200 GeV and for $M_N$= $10^4$ GeV respectively. It could be seen from these plots that larger the value of $y_\nu$, earlier $\lambda$ becomes negative and more is the tendency for the EW vacuum to be unstable as expected. We note from these two figures that for $\kappa$=0.304, absolute stability is attained only for $y_{\nu}$=0 even in the presence of the singlet scalar.

\begin{figure}
    \begin{subfigure}[b]{0.47\textwidth}
        \includegraphics[width=\textwidth]{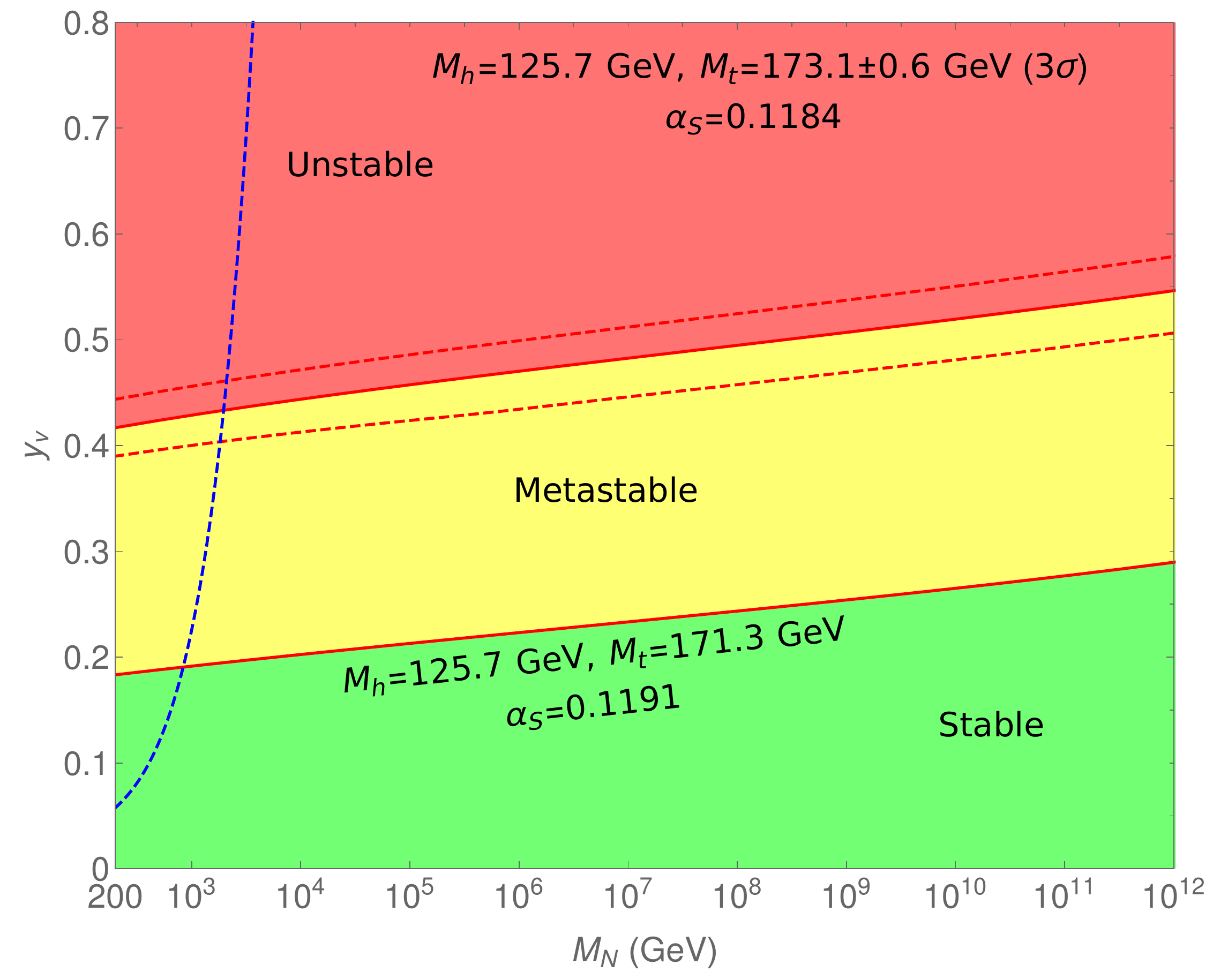}
        \caption{\centering Without the extra scalar}\label{5a}
         \end{subfigure}
       \begin{subfigure}[b]{0.47\textwidth}
        \includegraphics[width=\textwidth]{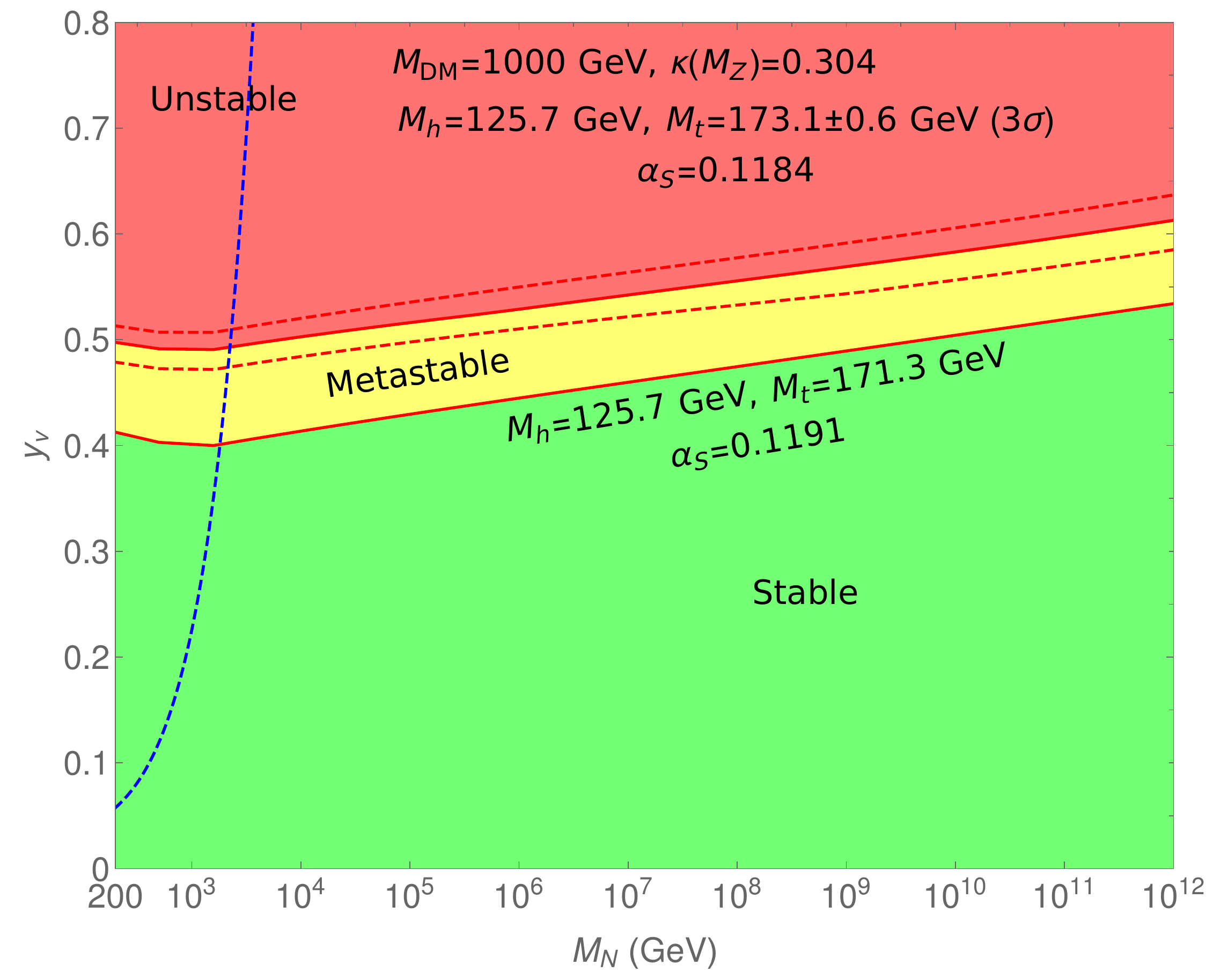}
        \caption{\centering With scalar, ($\kappa,\,M_{DM}$) = (0.304, 1000 GeV)}\label{5b}
                    \end{subfigure}
  \caption{Phase diagrams in the $y_\nu$ - $M_N$ plane in the presence and the absence of the extra scalar. Region in the left side of the blue dotted line is disallowed by 
constraint from BR($\mu \rightarrow e \gamma$). The three boundary lines (two dotted and a solid) 
correspond to $M_t \, = \, 173.1 \,\pm \, 0.6$ GeV (3$\sigma$) and we have taken $\lambda_S(M_Z) = 0.1$ in the second plot.}\label{linearynuMR}
\end{figure}

 In fig.(\ref{linearynuMR}), we have shown the phase diagram in the $y_\nu \,-\, M_N$ plane. The stable (green), unstable (red) and the metastable (yellow) regions are shown 
and it could be seen that higher the value of $M_N$, larger the allowed values of $y_\nu$ by vacuum stability as we have discussed earlier. The unstable and the metastable regions are separated by solid red line for the central values of the SM parameters, 
$M_h=125.7$ GeV, $M_{t}=173.1$ GeV and $\alpha_s=0.1184$. The red dashed lines represent the 3$\sigma$ variation of the top quark mass. However, we get significant stable region
for  $M_h=125.7$ GeV, $M_{t}=171.3$ GeV and $\alpha_s=0.1191$ which corresponds to the solid line separating the stable and the metastable region. The region in the left side 
of the blue dotted line is disallowed by LFV constraints for the normal hierarchy of light neutrino masses. Fig.(\ref{5a}) is drawn in the absence of the extra scalar and fig.(\ref{5b}) is drawn for ($\kappa,\,M_{DM}$) = (0.304, 1000 GeV). Clearly, there is more stable region in the presence of the extra scalar and the boundary line separating the metastable and the unstable regions also shifts upwards in this case.

\begin{figure}[h]
    \centering
    \begin{subfigure}[b]{0.45\textwidth}
        \includegraphics[width=\textwidth]{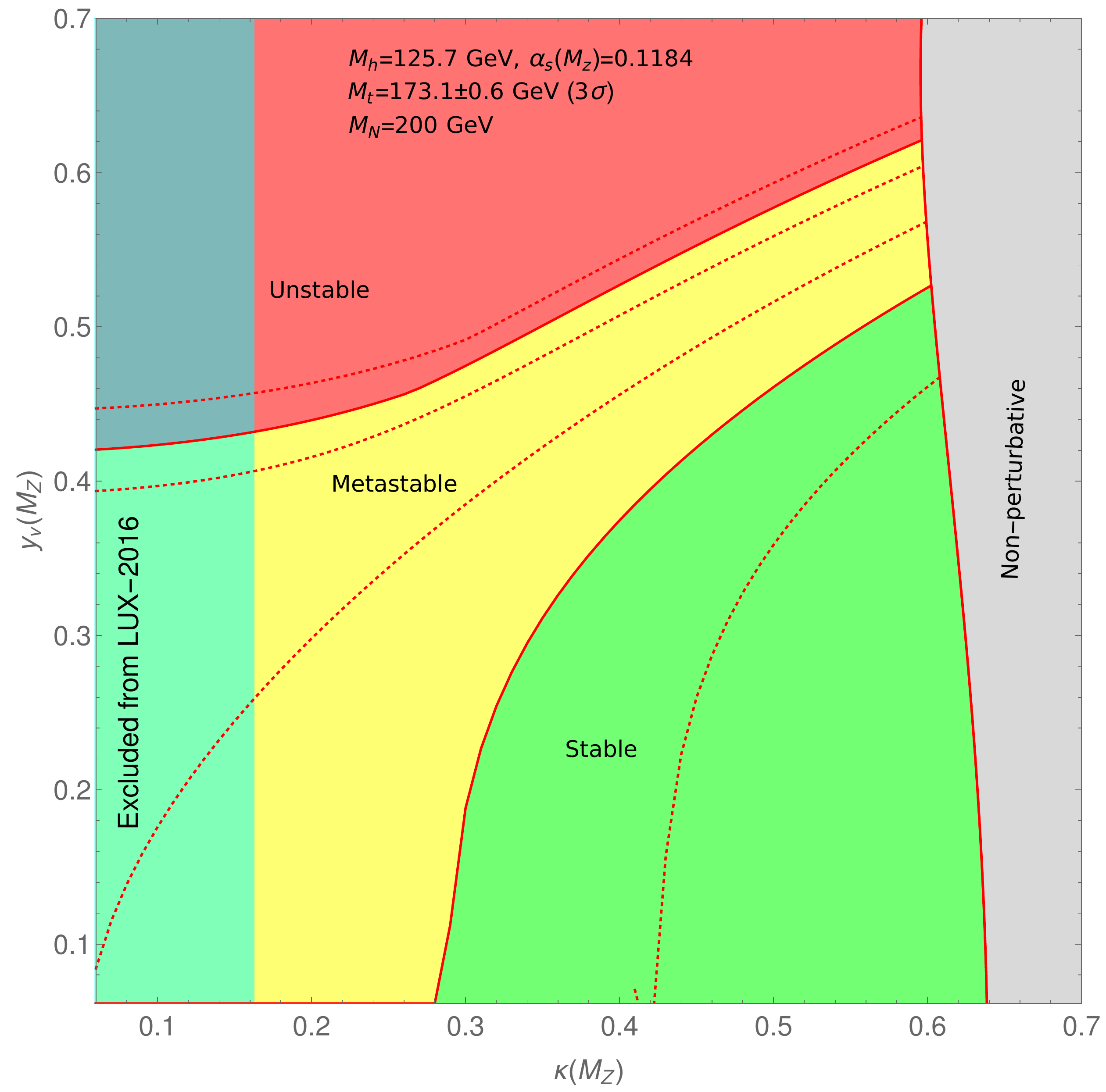}
        \caption{\centering }\label{6a}
         \end{subfigure}
       \begin{subfigure}[b]{0.45\textwidth}
        \includegraphics[width=\textwidth]{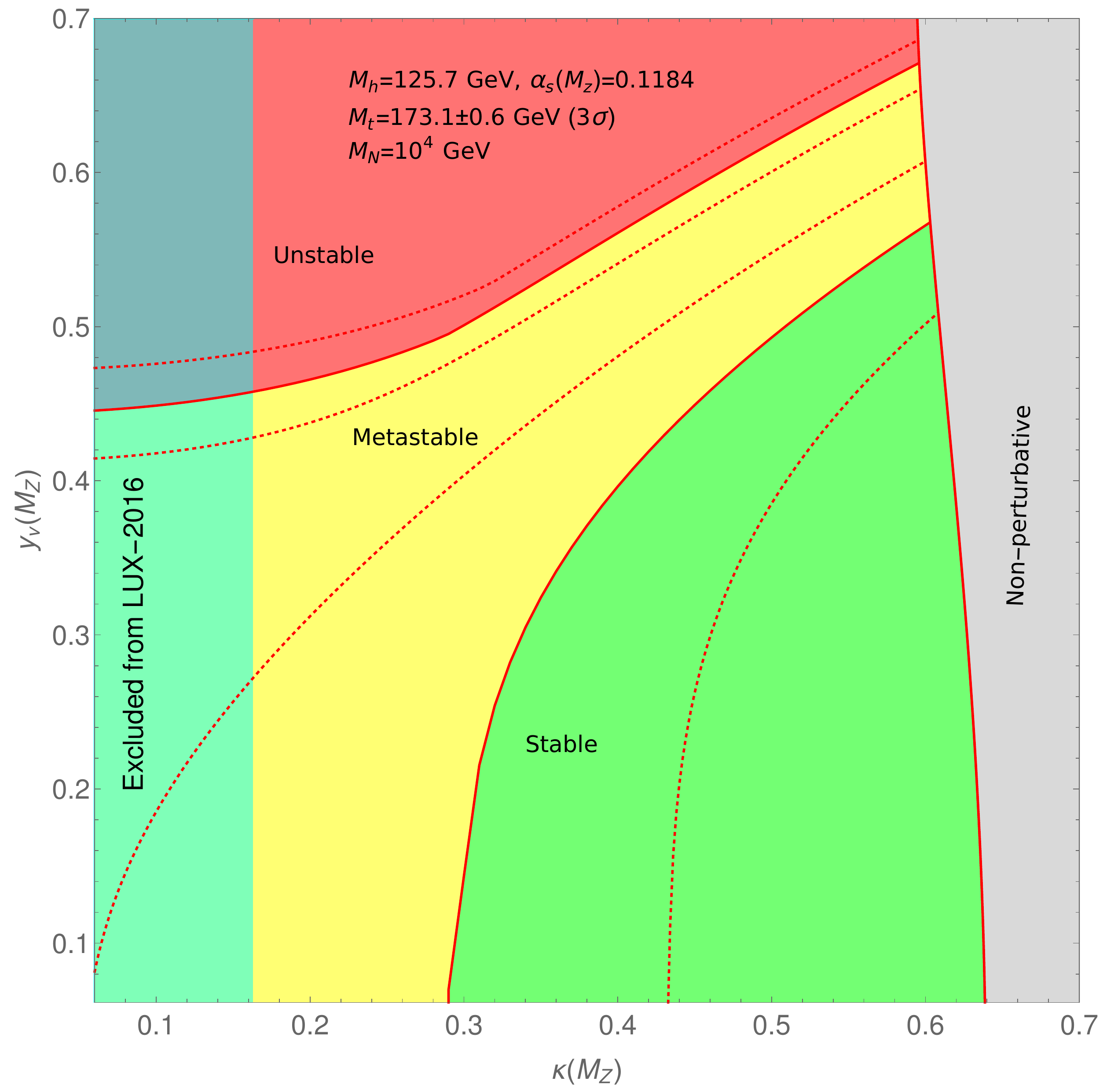}
        \caption{\centering }\label{6b}
                    \end{subfigure}
 
        \caption{\centering  Phase Diagrams in the $y_\nu$ - $\kappa$ plane for two different values of $M_N$. Here, $\lambda_S(M_Z)=0.1$.}\label{linearynukap}
\end{figure}

In fig.(\ref{linearynukap}), we have shown the phase diagrams in the $y_\nu$ - $\kappa$ plane for two different values of the heavy 
neutrino masses : fig.(\ref{6a}) for $M_N \,=\, 200 $ GeV and fig.(\ref{6b}) for $M_N \,=\, 10^4 $ GeV. Here also, the red dashed lines represent the 3$\sigma$ variation of top quark mass.  It could  clearly be 
seen that as the value of the heavy neutrino mass is higher, the unstable region shifts towards the large values of $y_{\nu}$. This is a result that should 
be expected from fig.(\ref{linearynuMR}).  In this model, the theory becomes non-perturbative (grey) for $\kappa$=0.64 for $y_{\nu}$=0.05. The maximum allowed value of $\kappa$ by perturbativity at the Planck scale decreases with increase in $y_{\nu}$ as we have also seen for the inverse seesaw case. The region $\kappa \lesssim 0.16$ is excluded from the recent direct detection experiment at LUX.

\section{Conclusions }
In this paper we have analysed the stability of the electroweak vacuum in the context of TeV scale inverse seesaw and minimal linear seesaw models extended with a scalar singlet dark matter. We have studied the interplay between the contribution of the extra  singlet scalar and the singlet fermions to the EW vacuum stability. We have shown that the coupling constants  in these two seemingly disconnected sectors can be correlated at high energy by the vacuum stability/metastability and perturbativity constraints. 

In the inverse seesaw scenario, the EW vacuum stability analysis is done after fitting the model parameters with the neutrino oscillation data and non-unitarity constraints on $U_{PMNS}$ (including  the LFV constraints   from $\mu \rightarrow e \gamma$). For the minimal linear seesaw model, the Yukawa matrix $Y_{\nu}$ can be fully parameterized in terms of  the oscillation parameters excepting an overall coupling constant $y_{\nu}$ which can be constrained  from vacuum stability and LFV.  We have taken the heavy neutrino masses of order upto a few TeV for both the seesaw models.  An extra $Z_2$ symmetry is imposed to ensure that the scalar particle
serves as a viable dark matter candidate. We include all the experimental and theoretical bounds  coming from the constraints on relic density and dark matter searches as well as unitarity and perturbativity upto the Planck scale. For the masses of new fermions from 200 GeV to a few TeV, the annihilation  cross section to the extra fermions is very small for dark matter mass ${\mathcal O}(1-2)$ TeV.   We have also checked that the theory violates perturbativity before the Planck scale for DM mass $\gtrsim 2.5 $ TeV. In addition we find that the value of the Higgs portal coupling $\kappa$ ($M_Z$) for which perturbativity is violated at the Planck scale decreases with increase in the value of the Yukawa couplings of the new fermions. For $M_{DM} >> M_{t}$, one can approximately write  $M_{DM}$ $\sim$ 3300 $\kappa$. This implies that with the increasing Yukawa coupling, the mass of dark matter for which the perturbativity is maintained also decreases. Thus the RGE running induces a correlation between the couplings of the two sectors from the perturbativity constraints.  

It is well known that the electroweak vacuum of SM is in the metastable region. The presence of the fermionic Yukawa couplings in the context of TeV scale seesaw models drives the vacuum more towards instability while the singlet scalar tries to arrest this tendency. Overall, we find that it is possible to find parameter spaces for which the electroweak vacuum remains absolutely stable for both inverse and linear seesaw models in the presence of the extra scalar particle.   We find an upper bound from metastability on Tr$[Y_\nu^\dag Y_\nu]$ as 0.25 for $\kappa$=0  which increases to 0.4 for $\kappa$=0.6 in inverse seesaw model. 
 We have also seen that in the absence of the extra scalar, the values of the Yukawa coupling $y_\nu $ greater than $0.42$ are disallowed in the minimal linear seesaw model. But, in the presence of the extra scalar the  values of  $y_\nu$ up to $\sim 0.6$ are allowed for dark matter mass  $\sim1$ TeV. The correlations between the Yukawa couplings (Tr$[Y_\nu^\dag Y_{\nu}]$ or $y_{\nu}$) and $\kappa$ are presented in terms of phase diagrams.
 
Inverse and linear seesaw models can be explored at LHC through trilepton signatures \cite{Bambhaniya:2014kga,delAguila:2008cj,Haba:2009sd,Chen:2011hc,Das:2012ze,Bandyopadhyay:2012px,Das:2014jxa,Das:2015toa,Mondal:2016kof}. A higher value of Yukawa couplings, as can be achieved in the presence of the Higgs portal dark matter, can facilitate observing such signals at colliders.

 \newpage
\bibliographystyle{unsrt}
\bibliography{tevportalnew}

\end{document}